\documentclass{article}

    \usepackage[nonatbib, preprint]{neurips_2020}

\usepackage[utf8]{inputenc} 
\usepackage[T1]{fontenc}    
\usepackage{hyperref}       
\usepackage{url}            
\usepackage{booktabs}       
\usepackage{amsfonts}       
\usepackage{nicefrac}       
\usepackage{microtype}      
\usepackage{xcolor}
\usepackage{graphicx}
\usepackage{booktabs} 
\usepackage{lipsum}
\usepackage{delimset}
\usepackage{amsthm}
\usepackage{amsfonts}  
\usepackage{algorithm}
\usepackage{algorithmic}
\usepackage[numbers]{natbib}
\usepackage{todonotes}
\usepackage{graphicx}
\usepackage{subcaption}
\usepackage{verbatim}

\newtheorem{theorem}{Theorem}
\newtheorem{lemma}[theorem]{Lemma}

\newcommand{\expib}{\rm{dualExpIB}}
\newcommand{\dualib}{\rm{dualIB}}
\newcommand{\vdib}{\rm{VdualIB}}
\newcommand{\vib}{\rm{VIB}}
\newcommand{\gvdib}{\rm{GVdualIB}}
\newcommand{\cvdib}{\rm{ConfVdualIB}}
\newcommand{\anvdib}{\rm{AnGVdualIB}}
\newcommand{\prdvib}{\rm{PrdTrVdualIB}}
\newcommand{\ib}{\rm{IB}}
\newcommand{\hX}{\hat{X}}
\newcommand{\hx}{\hat{x}}
\newcommand{\hY}{\hat{Y}}
\newcommand{\hy}{\hat{y}}

\newcommand{\rvx}{\boldsymbol{x}}
\newcommand{\rvy}{\boldsymbol{y}}
\newcommand{\lmref}[1]{\textit{lemma}~\ref{#1}}
\newcommand{\fref}[1]{Figure~\ref{#1}}

\title{The Dual Information Bottleneck}

\author{Zoe Piran\\
Racah Institute of Physics\\
The Hebrew University of Jerusalem\\
Jerusalem, Israel \\
\texttt{zoe.piran@mail.huji.ac.il} \\
\And
Ravid Shwartz-Ziv \\
School of Computer Science \\
The Hebrew University of Jerusalem\\
Jerusalem, Israel \\
\texttt{ravid.ziv@mail.huji.ac.il}
\And
Naftali Tishby \\
School of Computer Science\\
The Hebrew University  of Jerusalem\\
Jerusalem, Israel \\
\texttt{tishby@cs.huji.ac.il}
}

\begin{document}

\maketitle

\begin{abstract}
  The Information Bottleneck ({\ib}) framework is a general characterization of optimal representations obtained using a principled approach for balancing accuracy and complexity.
  Here we present a new framework, the Dual Information Bottleneck ({\dualib}), which resolves some of the known drawbacks of the {\ib}.
  We provide a theoretical analysis of the {\dualib} framework; (i) solving for the structure of its solutions (ii) unraveling its superiority in optimizing the \emph{mean prediction error exponent} and (iii) demonstrating its ability to preserve exponential forms of the original distribution.
  To approach large scale problems, we present a novel variational formulation of the {\dualib} for Deep Neural Networks. In experiments on several data-sets, we compare it to a variational form of the {\ib}. This exposes superior Information Plane properties of the {\dualib} and its potential in improvement of the error.
\end{abstract}

\section{Introduction}
\label{sec:intro}
The Information Bottleneck ({\ib}) method \cite{tishby99information}, is an information-theoretic framework for describing efficient representations of a ``input'' random variable $X$ that preserve the information on an ``output''  variable $Y$. In this setting the joint distribution of $X$ and $Y$, $p\brk*{x, y}$, defines the problem, or rule, and the training data are a finite sample from this distribution. The stochastic nature of the label is essential for the analytic regularity of the {\ib} problem. In the case of deterministic labels, we assume a {\emph{ noise model}} which induces a distribution. The representation variable $\hat{X}$  is in general a stochastic function of $X$ which forms a Markov chain $Y \rightarrow X \rightarrow \hat{X}$, and it depends on $Y$ only through the input $X$.   We call the map $p\brk*{\hx \mid x}$ the {\emph{encoder}} of the representation and denote by $p\brk*{y \mid \hx}$ the {\emph{Bayes optimal decoder}} for this representation; i.e., the best possible prediction of the \emph{desired label} $Y$ from the representation $\hX$.

The {\ib} has direct successful applications for representation learning in various domains, from vision and speech processing \cite{ma2019unpaired}, to neuroscience \cite{schneidman2001analyzing}, and Natural Language Processing \cite{li-eisner-2019}. 
Due to the notorious difficulty in estimating mutual information in high dimension, variational approximations to the {\ib} have been suggested and applied also to Deep Neural Networks (DNNs) \citep[e.g.,][]{Alemi2016DeepVI,Parbhoo2018CausalDI,poole2019variational}. Additionally, following \citep{shwartz2017}, several recent works tackled the problem of understanding DNNs using the {\ib} principle \citep{nash2018inverting, goldfeld2018estimating} 

Still, there are several drawbacks to the {\ib} framework which motivated this work. While the standard approach in representation learning is to
use the topology or a specific parametric  model over the input, the {\ib} is completely non-parametric and it operates only on the probability space. In addition, the {\ib} formulation does not relate to the task of prediction over unseen patterns and assumes full access to the joint probability of the patterns and labels.

Here, we resolve the above drawbacks by introducing a novel framework, the Dual Information Bottleneck ({\dualib}). The {\dualib} can account for 
known features of the data and use them to make better predictions over unseen examples, from small samples for large scale problems. Further, it emphasizes the prediction variable, $\hY$, which wasn't present in the original {\ib} formulation.

\subsection{Contributions of this work}
\label{sec:contribution}
We present here the Dual Information Bottleneck ({\dualib}) aiming to obtain optimal representations, which resolves the {\ib} drawbacks:
\begin{itemize}
\item We provide a theoretical analysis which obtains an analytical solution to the framework and compare its behaviour to the {\ib}.

\item For data which can be approximated by exponential families  we provide closed form solutions, {\expib}, which preserves the sufficient statistics of the original distribution.
\item We show that by accounting for the prediction variable, the {\dualib} formulation optimizes a bound over the error exponent. 

\item We present a novel variational form of the {\dualib} for Deep Neural Networks (DNNs) allowing its application to real world problems. Using it we empirically investigate the dynamics of the {\dualib} and validate the theoretical analysis.
\end{itemize}

\section{Background} \label{sec:background}

The Information Bottleneck ({\ib}) framework is defined as the trade off between the encoder and decoder mutual information values. It is defined by the minimization of the Lagrangian:
\begin{align}
\label{eq:IB_L}
    \mathcal{F}\brk[s]*{p_{\beta}\brk*{\hat{x} \mid x}; p_{\beta}\brk*{y \mid \hat{x} }}  =I(X;\hat{X}) - \beta I(Y; \hat{X})~,
\end{align}

independently over the convex sets of the normalized distributions, $\brk[c]*{p_{\beta}\brk*{\hat{x} \mid x}}$, $\brk[c]*{p_{\beta}\brk*{\hat{x}}}$ and $\brk[c]*{p_{\beta}\brk*{y \mid \hat{x}}}$, given a positive Lagrange multiplier $\beta$ constraining the information on $Y$, while preserving the Markov Chain $Y \rightarrow X \rightarrow \hat{X}$. Three self-consistent equations for the optimal encoder-decoder pairs, known as the \emph{{\ib} equations}, define the solutions to the problem. An important characteristic of the equations is the existence of critical points along the optimal line of solutions in the \emph{information plane} (presenting $I(Y;\hX)$ vs. $I(X;\hX)$) \citep{wu2020phase, parker}. 
The {\ib} optimization trade off can be considered as a generalized rate-distortion problem \cite{Cover:2006:EIT:1146355} with the distortion function, $d_{\ib}\brk*{x,\hx}=D\brk[s]*{p\brk*{y \mid x}||p_{\beta}\brk*{y|\hx}}$.
For more background on the {\ib} framework see \S \ref{app:ib}. 

\section{The Dual Information Bottleneck}
\label{sec:dualIB}

Supervised learning is generally separated into two phases: the training phase, in which 
 the internal representations are formed from the training data, and the prediction phase, in which  these representations are used to predict labels of new input patterns  \citep{shalev2014understanding}.
To explicitly address these different phases we add to the {\ib} Markov chain another variable, $\hY$, the \emph{predicted label} from the trained representation, which obtains the same values as $Y$ but is distributed differently: 
\begin{align} \label{eq:MC}
\rlap{$\overbrace{\phantom{Y \rightarrow X \rightarrow \hat{X}_{\beta}}}^{\textrm{training}}$}Y \rightarrow \underbrace{X \rightarrow
     \hat{X}_{\beta}\rightarrow \hat{Y}}_{\textrm{prediction}}.
\end{align}
The left-hand part of this chain describes the representation training, while the right-hand part is the Maximum Likelihood prediction using these representations \cite{slonim_MIB}. So far the prediction variable $\hY$ has not been a part of the $\ib$ optimization problem. 
It has been implicitly assumed that the \emph{Bayes optimal decoder}, $p_{\beta}\brk*{y \mid \hx}$, which maximizes the full representation-label information, $I(Y;\hX)$, for a given $\beta$, is also the best choice for making predictions. Namely, the prediction of the label, $\hY$, from the representation $\hX_{\beta}$ through the right-hand Markov chain by the mixture using the internal representations,
$ p_{\beta}\brk*{\hat{y} \mid x} \equiv \sum_{\hat{x}} p_{\beta}\brk*{y= \hat{y} \mid \hat{x}}  p_{\beta}\brk*{\hat{x} \mid x},$
is optimal when $p_{\beta}\brk*{y \mid \hx}$ is the \emph{Bayes optimal decoder}. However, this is not necessarily the case, for example when we train from finite samples \cite{DBLP:conf/alt/ShamirST08}.
 
 Focusing on the prediction problem, we define the {\dualib} distortion by switching the order of the arguments in the KL-divergence of the original {\ib} distortion, namely:
\begin{align}\label{eq:dist}
d_{\dualib}\brk*{x,\hx}&=D\brk[s]*{p_{\beta}\brk*{y\mid \hx} \| p\brk*{y \mid x}} 
=\sum_{y}p_{\beta}\brk*{y\mid \hx}\log \frac{p_{\beta}\brk*{y\mid \hx}}{p\brk*{y \mid x}}\ .    
\end{align}
In geometric terms this is known as the \emph{dual} distortion problem \cite{Ay2019}. The $\dualib$ optimization can then be written as the following rate-distortion problem:
\begin{align} \label{eq:min_func_dual_ib}
    \mathcal{F}^{*}\brk[s]*{p_{\beta}\brk*{\hat{x} \mid x} ;p_{\beta}\brk*{y \mid \hx }} &=  I(X;\hat{X})  
    + \beta \mathbb{E}_{p_{\beta}\brk*{x, \hx}}\brk[s]*{d_{\dualib}\brk*{x, \hx} }~.
\end{align}
As the decoder defines the prediction stage ($p_{\beta}\brk*{y=\hy \mid \hx}$) we can write (see proof in \S \ref{app:dual_ib}) the average distortion in terms of mutual information on $\hY$,
$I(\hX ; \hY)$ and $I(X ;\hY)$:
\begin{align} \label{eq:dual_dist_MI}
       \mathbb{E}_{p_{\beta}\brk*{x, \hx}}\brk[s]*{d_{\dualib}\brk*{x, \hx}} &=   \underbrace{I(\hX;\hY) - I( X;\hY)}_{(a)} 
       +  \underbrace{ \mathbb{E}_{p\brk*{x}}\brk[s]*{D\brk[s]*{p_{\beta}\brk*{\hy \mid x} \| p\brk*{y=\hy \mid x} }}}_{(b)} 
.\end{align}
This is similar to the known {\ib} relation: $ \mathbb{E}_{p_{\beta}\brk*{x, \hx}}\brk[s]*{d_{\ib}\brk*{x, \hx}} = I(Y;X) - I(Y;\hX) $
with an extra positive term $(b)$. Both terms, $(a)$ and $(b)$, vanish precisely when $\hX$ is a sufficient statistic for $X$ with respect to $\hY$. In such a case we can reverse the order of $X$ and $\hX$ in the Markov chain \eqref{eq:MC}. This replaces the roles of $Y$ and $\hY$ as the variable for which the representations, $\hX_{\beta}$, are approximately minimally sufficient statistics. 
In that sense the {\dualib}  shifts the emphasis from the training phase to the prediction phase.  This implies that minimizing the {\dualib} functional maximizes a lower bound on the mutual information $I(X;\hY)$.

\subsection{The {\dualib} equations}
\label{sec:dualIB_sols}
Solving the {\dualib} minimization problem \eqref{eq:min_func_dual_ib}, we obtain a set of self consistent equations. Generalized Blahut-Arimoto (BA) iterations between them converges to an optimal solution. The equations are similar to the original {\ib} equations with the following modifications: (i) Replacing the distortion by its dual in the encoder update; (ii) Updating the decoder by the encoder's geometric mean of the data distributions $p\brk*{y \mid x}$.
\begin{theorem}
\label{lm:dual_ib_eqs}
The {\dualib} equations are given by:
\begin{align} \label{eq:dIB}
	\begin{cases}
	\brk*{i}\ &p_{\beta}\brk*{\hx \mid x} = \frac{p_{\beta}\brk*{\hx} }{Z_{\hat{\rvx} \mid \rvx}\brk*{x;\beta} } e^{-\beta D\brk[s]*{ p_{\beta} \brk*{y \mid \hx} \| p\brk*{y\mid x} }} \\
	\brk*{ii}\ &p_{\beta}\brk*{\hat{x}} = \sum_{x} p_{\beta}\brk*{\hat{x} \mid x} p\brk*{x} \\
	\brk*{iii}\ &p_{\beta}\brk*{y \mid \hx } = \frac{ 1}{Z_{\rvy\mid \hat{\rvx}}\brk*{\hx; \beta}} \prod_{x}   p\brk*{y \mid x}^{p_{\beta}\brk*{x \mid \hx}} 
	\end{cases}
,\end{align}
where $Z_{\hat{\rvx} \mid \rvx}\brk*{x;\beta}, Z_{\rvy\mid \hat{\rvx}}\brk*{\hx; \beta}$ are normalization terms.
\end{theorem}
The proof is given in \S \ref{app:dual_ib_eq_proof}. 
It is evident that the basic structure of the equations of the {\dualib} and {\ib} is similar and 
they approach each other for large values of $\beta$. In the following sections we explore the implication of the differences on the properties of the new framework.

\subsection{The critical points of the {\dualib}}
\label{sec:bifurcation_points_dual}

As mentioned in \S \ref{sec:background} and \cite{wu2020phase}, the ``skeleton'' of the {\ib} optimal bound (the information curve) is constituted by the critical points in which the topology (cardinality) of the representation changes. 
Using perturbation analysis over the  {\dualib} optimal representations we find that small changes in the encoder and decoder that satisfy \eqref{eq:dIB} for a given $\beta$ are approximately determined through a nonlinear eigenvalues problem. \footnote{For simplicity we ignore here the possible interactions between the different representations.}

\begin{theorem} \label{th:stab_anl_dual}
The {\dualib} critical points are given by non-trivial solutions of the nonlinear eigenvalue problem:
\begin{align} \label{eq:stab_anl_dual}
	\brk[s]*{I - \beta 	C^{\dualib}_{xx'} \brk*{\hat{x}, \beta }} {\delta \log p_{\beta}\brk*{x' \mid \hat{x}}}&=0  
	~,~~~~  
	\brk[s]*{I - \beta C^{\dualib}_{y y'}\brk*{\hat{x}, \beta }}{\delta \log p_{\beta}\brk*{y' \mid \hat{x}}}=0
.\end{align}
The matrices $C^{\dualib}_{xx'}\brk*{\hat{x}; \beta}, C^{\dualib}_{yy'}\brk*{\hat{x}; \beta}$ have the same eigenvalues  $\brk[c]*{\lambda_{i}}$, with $\lambda_{1}(\hx) = 0$. With binary $y$, the critical points are obtained at $\lambda_{2}\brk*{\hx} =\beta^{-1}$.
\end{theorem}

\begin{figure}[htb!]
    \begin{center}
        \begin{subfigure}{.3\textwidth}
    \centering
    \includegraphics[width=\textwidth]{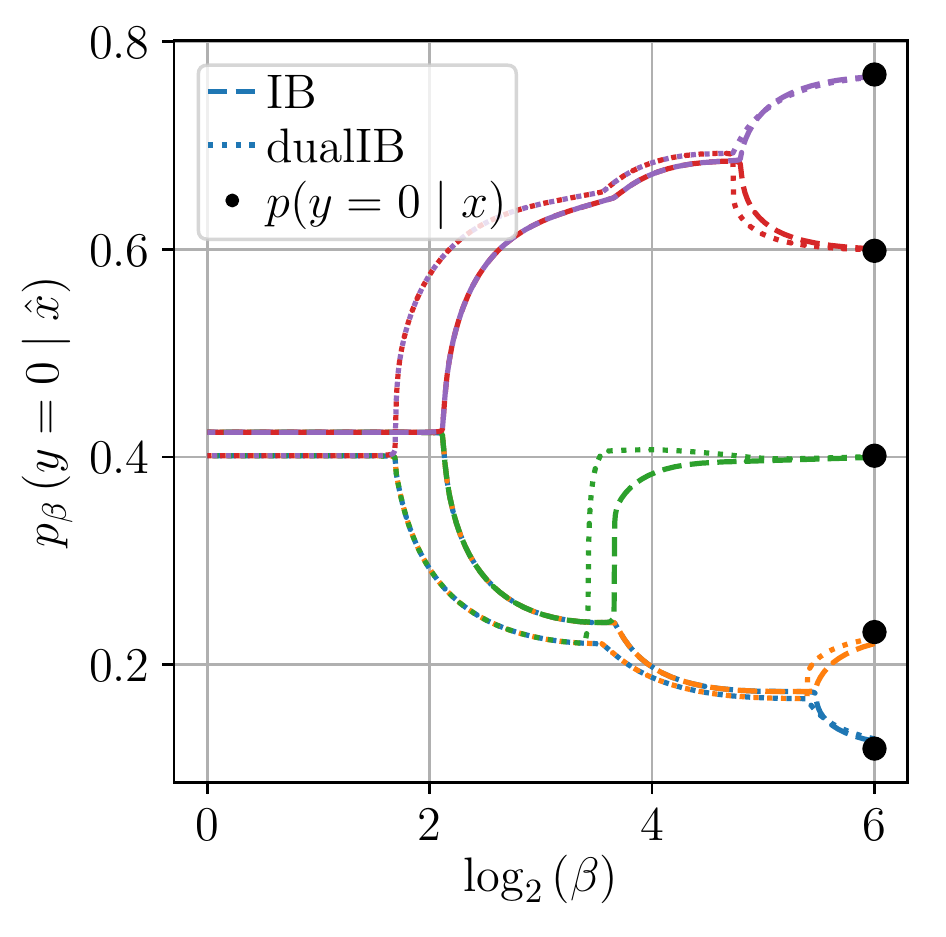}
    \caption{The bifurcation diagram}
    \label{fig:bif_a}
\end{subfigure}
            \begin{subfigure}{.3\textwidth}
    \centering
    \includegraphics[width=\textwidth]{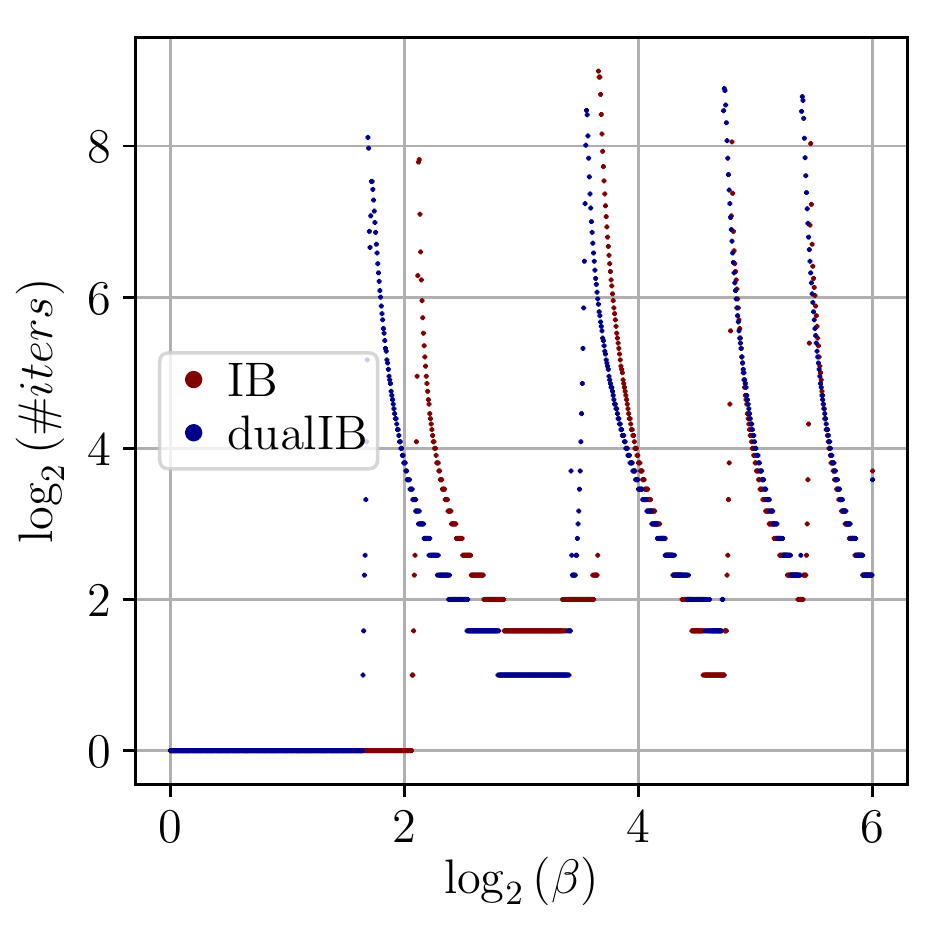}
    \caption{Convergence Time}
    \label{fig:bif_b}
\end{subfigure}
        \begin{subfigure}{.3\textwidth}
    \centering
    \includegraphics[width=\textwidth]{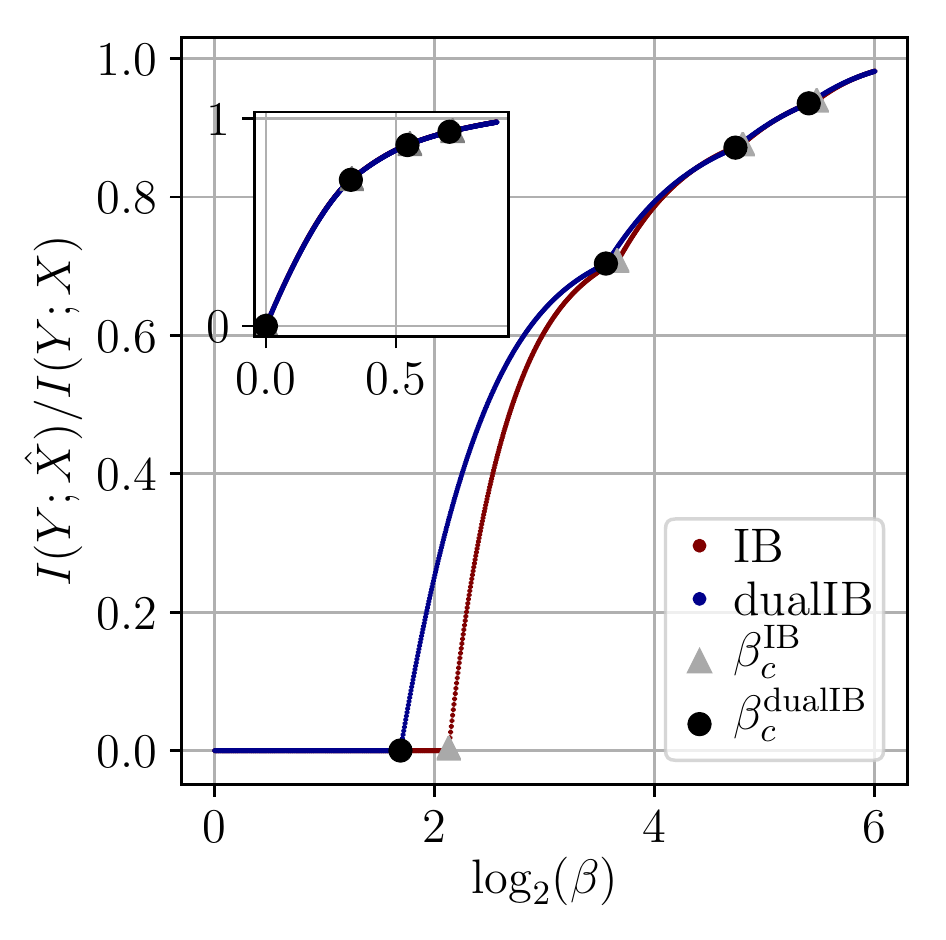}
    \caption{The information plane}
    \label{fig:bif_c}
\end{subfigure}
        \caption{$\brk{a}$ The \emph{bifurcation diagram}; each color corresponds to one component of the representation $\hx \in \hX$ and depicts the decoder $p_{\beta}\brk*{y = 0\mid \hat{x}}$. Dashed lines represent the {\ib}'s solution and dotted present the {\dualib}'s. The black dots denote the input distribution $p\brk*{y = 0 \mid x}$. $\brk*{b}$ Convergence time the BA algorithms as a function of $\beta$. $\brk*{c}$
         The desired label Information $I^{\ib}_{y}\brk*{\beta}$ and $I^{\dualib}_{y}\brk*{\beta}$ as functions of $\beta$. The inset shows the information plane, $I_{y}$  vs. $I_{x}$. The black dots are the {\dualib}'s critical points, $\beta^{\dualib}_c$, and the grey triangles are the {\ib}'s, $\beta^{\ib}_c$}
        \label{fig:bif_comb}
        \end{center}
  \end{figure}

The proof to Theorem \ref{th:stab_anl_dual} along with the structure of the matrices  $C^{\dualib}_{xx'}\brk*{\hat{x}; \beta}, C^{\dualib}_{yy'}\brk*{\hat{x}; \beta}$ is given in \S \ref{app:stability_anl}.
We found that this solution is similar to the nonlinear eigenvalues problem for the {\ib}, given in \S \ref{app:ib}.
As in the {\ib}, at the critical points we observe cusps 
with an undefined second derivative in the mutual information values as functions of $\beta$ along the optimal line.
That is the general structure of the solutions is preserved between the frameworks, as can be seen in \fref{fig:bif_c}. 

The \emph{Information Plane}, $I_{y}=I(\hX;Y)$  vs. $I_{x}=I(\hX;X)$, is the standard depiction of the compression-prediction trade-off of the {\ib} and has known analytic properties\citep{Gilad-bachrach}. First, we note that both curves obey similar constraints, as given in \lmref{lm:info_concave} below. 
\begin{lemma} \label{lm:info_concave}
  along the optimal lines of $I_{x}(\beta)$ and $I_{y}(\beta)$ the curves are non-decreasing
 piece-wise concave 
 functions of $\beta$. 
 When their second derivative (with respect to $\beta$) is defined, it is strictly negative.
\end{lemma} 

Next, comparing the {\dualib}'s and {\ib}'s information planes we find several interesting properties which are summarized in the following {Theorem} (see \S \ref{app:performance_anl} for the proof).

\begin{theorem} \label{th:dib_plane}
(i) The critical points of the two algorithms alternate, $\forall i, i+1$, $\beta_{c,i}^{\dualib} \leq \beta_{c,i}^{\ib} \leq \beta_{c,i+1}^{\dualib} \leq \beta_{c,i}^{\ib}$. (ii) The  distance between the two information curves is minimized at $\beta_{c}^{\dualib}$. (iii) The two curves approach each other as $\beta \rightarrow \infty$.
\end{theorem}

From Theorem \ref{th:dib_plane} we deduce that as the dimensionality of the problem increases (implying the number of critical points grows) the {\dualib}'s approximation of the {\ib}'s information plane becomes tighter. We illustrate the behavior of the  {\dualib}'s solutions in comparison to the {\ib}'s on a low-dimensional problem that is easy to analyze and visualize, with a binary $Y$ and only $5$ possible inputs $X$ (the complete definition is given in \S \ref{app:def_prob}). For any given $\beta$, the encoder-decoder iterations converge to stationary solutions of the \emph{{\dualib}} or \emph{{\ib} equations}. The evolution of the optimal decoder, $p_{\beta}\brk*{y=0 \mid \hx}$, $\forall \hx \in \hX$, as a function of $\beta$, forms a \emph{bifurcation diagram} (\fref{fig:bif_a}), 
 in which 
the critical points define the location of the bifurcation. At the critical points the number of iterations diverges (\fref{fig:bif_b}).
While the overall structure of the solutions is  similar, we see a ``shift'' in the appearance of the representation splits between the two frameworks. Specifically, as predicted by Theorem \ref{th:dib_plane} the {\dualib} bifurcations occur at lower $\beta$ values than  those of the {\ib}. The inset of \fref{fig:bif_c} depicts this comparison between the two information curves. While we know that $I^{\ib}_{y}\brk*{\beta}$ is always larger, we see that for this setting the two curves are almost indistinguishable.  Looking at $I_{y}$ as a function of  $\beta$ (\fref{fig:bif_c}) the importance  of the critical points is revealed as the corresponding cusps along these curves correspond to ``jumps'' in the accessible information. Furthermore, the distance between the curves is minimized precisely at the dual critical points, as predicted by the theory. 

\section{The Exponential Family dualIB}
\label{sec:dualExpIb}

One of the major drawbacks of the {\ib} is that it fails to capture an existing parameterization of the data, that act as minimal sufficient statistics for it.
Exponential families are the class of parametric distributions for which minimal sufficient statistics exist,  forming an elegant theoretical core of parametric statistics and often emerge as maximum entropy \cite{Jaynes58} or stochastic equilibrium distributions, subject to observed expectation constraints. 
As the {\ib} ignores the structure of the distribution, given data from an exponential family it won't consider these known features. Contrarily, the {\dualib} accounts for this structure and its solution are given in terms of these features, defining the {\expib} equations.

We consider the case in which the rule distribution is of the form, $p\brk*{y \mid x} = e^{-\sum_{r=0}^{d} \lambda^r(y)A_r(x)} $,
where $A_r(x)$ are $d$ functions of the input $x$  and $\lambda^r(y)$ are functions of the label $y$, or the parameters of this exponential family \footnote{
The normalization factors, $Z_{\rvy \mid \rvx}\brk*{x}$, are written, for brevity, as $\lambda_{\rvx}^{0} \equiv \log ( \sum_{y} \prod_{r=1}^{d} e^{- {\lambda}^{r}\brk*{y} A_{r} \brk*{x} })$ with $A_{0} \brk*{x} \equiv 1 $.
We do not constrain the marginal $p\brk*{x}$.}. For exponential forms the mutual information, $I(X;Y)$, is fully captured by the $d$ conditional expectations. This implies that all the relevant information (in the training sample) is captured by $d$-dimensional empirical expectations which can lead to a reduction in computational complexity.

Next we show that for the {\dualib}, for all values of $\beta$, this dimension reduction is preserved or improved along the dual information curve. 
The complete derivations are given in \S \ref{app:dualexpIB}.

\begin{theorem} ({\expib})
\label{th:Exp-reduction}
For data from an exponential family the optimal encoder-decoder of the {\dualib} are given by:
\begin{align}  
\label{eq:dual_dec}
	 p_{\beta}\brk*{\hx\mid x} &= \frac{p_{\beta}\brk*{\hx}e^{\beta {\lambda}^{0}_{\beta}\brk*{\hx}}}{Z_{\hat{\rvx}\mid \rvx}\brk*{x;\beta} }  e^{-\beta  \sum_{r=1}^{d} {\lambda}_{\beta}^r\brk*{\hat{x}}   \brk[s]*{A_r \brk*{x} - A_{r, \beta}\brk*{\hat{x}}} } \nonumber \\
p_{\beta}\brk*{y\mid  \hx }  &= e^{-\sum_{r=1}^d \lambda^r \brk*{y} A_{r,\beta} \brk*{\hat{x}} -\lambda^0_{\beta}(\hx)} 
~, ~~~
 \lambda^0_{\beta}(\hx) = \log (\sum_{y} e^{-\sum_{r=1}^{d}  \lambda^r\brk*{y} A_{r,\beta} \brk*{\hx}})
,\end{align}
 with the constraints and multipliers expectations,
\begin{align}  
A_{r,\beta}\brk*{\hat{x}} &\equiv \sum_{x} p_{\beta}\brk*{x\mid \hat{x}} A_r \brk*{x}  
~,~
\lambda^r_{\beta}\brk*{\hat{x}} \equiv \sum_{y} p_{\beta}\brk*{y\mid \hat{x}} \lambda^r\brk*{y}~,~ 1\le r \le d ~.
\end{align}
\end{theorem}

This defines a simplified iterative algorithm  for solving
the {\expib} problem. Given the mapping of $x \in X$ to $\brk[c]*{A_{r}\brk*{x}}_{r=1}^{d}$ the problem is completely independent of $x$ and we can work in the lower dimensional embedding of the features, $A_{r}\brk*{x}$. 
Namely,   the update procedure 
is reduced to
 the dimensions of the sufficient statistics. Moreover, the representation is given in terms of the original features, a desirable feature for any model based problem.

\section{Optimizing the error exponent}

\label{sec:min_err_exp}
The {\dualib} optimizes an upper bound on the error exponent of the representation multi class testing problem. The error exponent accounts for the decay of the prediction error as a function of data size $n$. This implies the {\dualib} tends to  minimize the prediction error. For the classical binary hypothesis testing, the classification Bayes error, $P^{(n)}_e$, is the weighted sum of type 1 and type 2 errors. For large $n$, both errors decay exponentially with the test size $n$, and the best error exponent, $D^{*}$, is given by the Chernoff information. The Chernoff information is also a measure of distance  defined as, $ C\brk*{p_0, p_1}= \min_{0<\lambda <1 } \brk[c]*{\log \sum_{x} p_0^{\lambda}\brk*{x}p_1^{1-\lambda}\brk*{x}} $,  
 and we can understand it as an optimization on the $\log$-partition function of $p_{\lambda}$ to obtain $\lambda$ (for further information see \cite{Cover:2006:EIT:1146355} and \S \ref{app:err_exp}). 

The intuition behind the optimization of 
$D^{*}$ by the {\dualib} is in its distortion, the order of the prediction and the observation which implies the use of geometrical mean. The best achievable exponent (see \cite{Cover:2006:EIT:1146355}) in Bayesian probability of error is given by the KL-distortion between $p_{\lambda^{*}}$ ($ \propto p_0^{\lambda^*}\brk*{x}p_1^{1-\lambda^*}\brk*{x}$) to $p_0$ or $p_1$, such that $p_{\lambda^*}$ is the mid point between $p_0$ and $p_1$ on the geodesic of their geometric means.  
 By mapping the {\dualib} decoder to $\lambda$, it follows that the above minimization is proportional to the $\log$-partition function of $p_{\beta}\brk*{x \mid \hx}$, namely 
we obtain the mapping $p_{\beta}\brk*{x \mid \hx} = p_{\lambda}$. 

In the generalization to 
 multi class classification 
the error exponent is given by the pair of hypotheses with the minimal Chernoff information \cite{westover2008asymptotic}.
However, finding this value is generally 
difficult 
as it requires solving for each pair in the given classes.  Thus,
we consider an upper bound to it, the mean of the Chernoff information terms over classes. The representation variable adds a new dimension on which we average on and we obtain a bound on the optimal (in expectation over $\hx$) achievable exponent, $\hat{D}_{\beta}
 =\min_{p_{\beta}\brk*{y\mid \hx}, p_{\beta}\brk*{ \hx \mid x}}\mathbb{E}_{p_{\beta}\brk*{y, \hx}}\brk[s]*{D\brk[s]*{ p_{\beta}\brk*{ x \mid \hx} \mid p\brk*{x\mid y}}}$.
This expression is bounded from above by the {\dualib} minimization problem. 
Thus, the {\dualib} (on expectation) minimizes the prediction error for every $n$. A formal derivation of the above along with an analytical example of a multi class classification problem is given in \S \ref{app:err_exp}. In \S \ref{sec:experiment_setsize} we experimentally demonstrate that this also holds  for a variational {\dualib} framework using a DNN.

\section{Variational Dual Information Bottleneck}
The Variational Information Bottleneck (VIB) approach introduced by Alemi et al. \citep{Alemi2016DeepVI} allows to parameterize the {\ib} model using Deep Neural Networks (DNNs). The variational bound of the {\ib} is obtained using DNNs for both the encoder and decoder. 
Since then, various extensions have been made \citep{strouse2017deterministic, elad2019direct} demonstrating promising attributes. 
Recently, along this line, the Conditional Entropy Bottleneck (CEB) \citep{fischer2018conditional} was proposed. The CEB 
provides variational optimizing bounds on $I(Y;\hX)$, $I(X;\hX)$ using  a variational decoder $q\brk*{y\mid \hx}$, variational conditional marginal, $q\brk*{\hx\mid y}$, and a variational encoder,  $p\brk*{\hx\mid x}$, all implemented by DNNs.

Here, we present the variational {\dualib} ({\vdib}), which optimizes the variational {\dualib} objective for using in DNNs. 
Following the CEB formalism, we bound the {\dualib} objective. We develop a variational form of the {\dualib} distortion and combine it with the bound for $I(X;\hX)$ (as in the CEB). This gives us the following Theorem (for the proof see \S \ref{app:vdib_obj}.).
\begin{theorem}
The {\vdib} objective is given by:
\begin{align}
    \min_{q\brk*{\hx \mid y}, p\brk*{\hx\mid x}}\brk[c]*{\mathbb{E}_{\tilde{p}\brk*{y\mid x}p\brk*{\hx\mid x}p\brk*{x}}\brk[s]*{
   \log\frac{p\brk*{\hx\mid x}}{q\brk*{\hx\mid y}} } + \beta \mathbb{E}_{p\brk*{y\mid\hx}p\brk*{\hx\mid x}}\brk[s]*{\log \frac{p\brk*{y\mid \hx}}{\tilde{p}\brk*{y\mid x}}}}
,\end{align}
where $\tilde{p}\brk*{y \mid x}$ is a distribution based on the given labels of the data-set, which we relate to as the noise model. Under the assumption that the noise model captures the distribution of the data the above provides a variational upper bound to the {\dualib} functional \eqref{eq:min_func_dual_ib}.
\end{theorem}
Due to the nature of its objective the {\dualib} requires a noise model.
 We must account for the contribution to the objective arising
 from $\tilde{p}\brk*{y \mid x}$
 which could be ignored in the {\vib} case. The noise model can be specified by its assumptions over the data-set. In  \S \ref{sec:cifar10_inf} we elaborate on possible noise models choices and their implications on the performance. Notice that the introduction of $\tilde{p}(y \mid x)$ implies that, unlike most machine learning models, the {\vdib} does not optimize directly the error between the predicted and desired labels in the training data. Instead, it does so implicitly with respect to the noisy training examples. This is not unique to the {\vdib}, as it is equivalent to training with noisy labels, often done to prevent over-fitting. For example, in \citep{muller2019does} the authors show that label noise can improve  generalization that results in a reduction in the mutual information between the input and the output. 

In practice, similarly to the CEB, for the stochastic encoder,  $p(\hx \mid x)$, we use the original architecture, replacing the final softmax layer with a linear dense layer with $d$ outputs. These outputs are taken as the means of a multivariate Gaussian distribution with unit diagonal covariance. For the variational decoder, $q(y \mid \hx)$, any classifier network can be used. We take a linear softmax classifier which takes the encoder as its input. The reverse decoder $q(\hx \mid y)$ is implemented by a network which maps a one-hot representation of the labels to the $d$-dimensional output interpreted as the mean of the corresponding Gaussian marginal.

\subsection{Experiments}
To investigate the {\vdib} on real-world data we compare it to the CEB model 
using a DNN over two data-sets, FasionMNIST and CIFAR10. 
For both, we use a Wide ResNet $28-10$ \citep{zagoruyko2016wide} as the encoder, a one layer Gaussian decoder and a single layer linear network for the reverse decoder (similarly to the setup 
in \citep{fischer2020ceb}). We use the same architecture to  train networks with {\vdib} and {\vib} objectives.
(See \S \ref{app:exp_setup} for details on the experimental setup). We note that in our attempts to train over the CIFAR100 data-set the results did not fully agree with the results on the above data-sets (for more information see \S \ref{app:cifar100}).
An open source implementation is available \href{https://github.com/ravidziv/dual_IB.git}{here}. 

\subsubsection{The variational information plane}
As mentioned, the information plane describes the compression-prediction trade-off. It enables us to compare 
different models and evaluate their ``best prediction level'' in terms of the desired label information, for each compression level. 
In \citep{fischer2020ceb} the authors provide empirical evidence that information bottlenecking techniques can improve both generalization and robustness. Other works \citep{fischer2018conditional, achille2018emergence, achille2018information} provide both theoretical and conceptual  insights into why these improvements occur.

\begin{figure}
\begin{subfigure}{.3\textwidth}
    \centering
    \includegraphics[width=\textwidth]{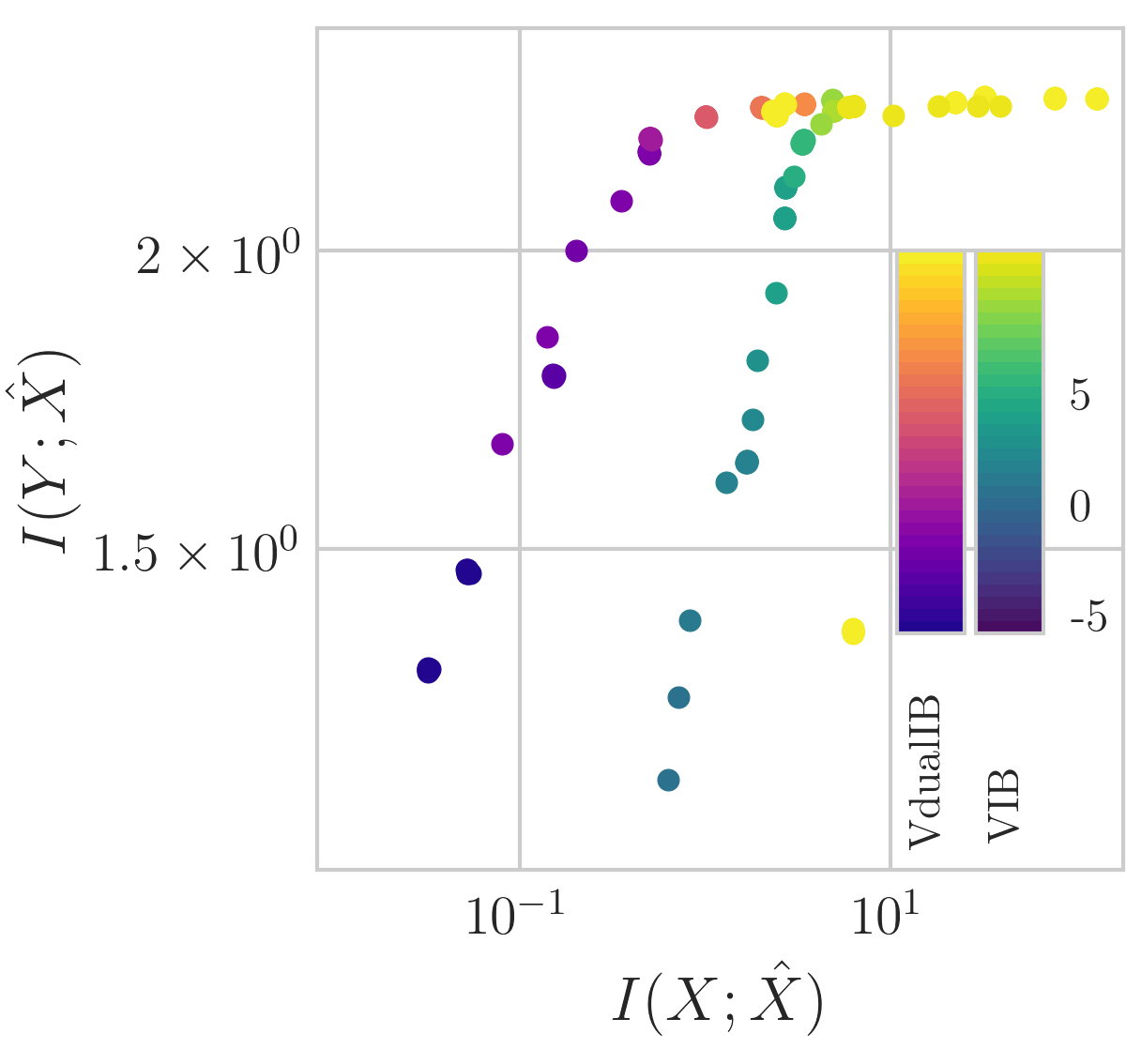}
    \caption{The information plane}
    \label{fig:inf_plane_mnist_beta}
\end{subfigure}
\begin{subfigure}{.3\textwidth}
    \centering
    \includegraphics[width=\textwidth]{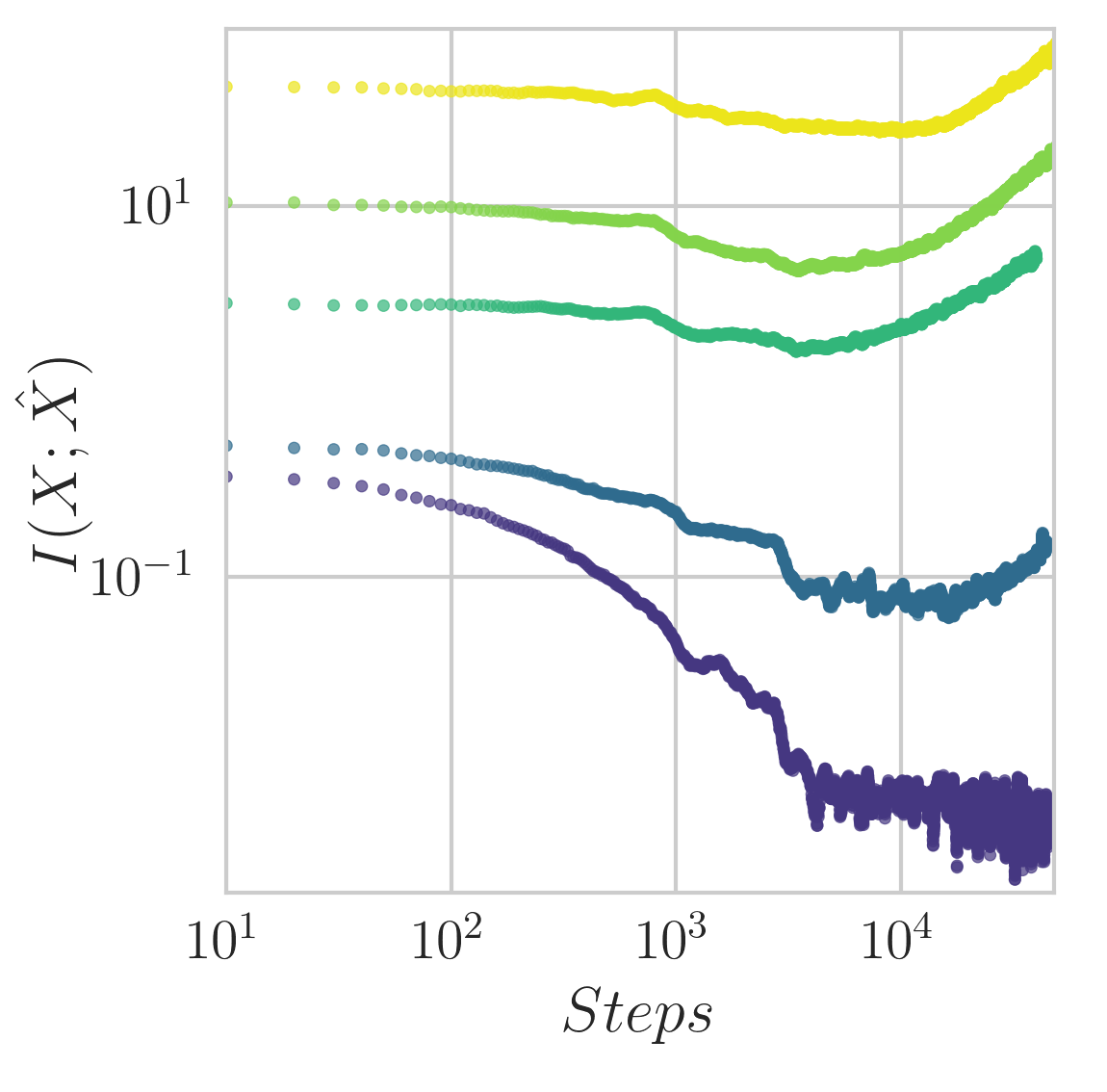}
    \caption{$I(X;\hX)$ vs. update steps}
    \label{fig:inf_plane_mnist_compare}
\end{subfigure}
\begin{subfigure}{.3\textwidth}
    \centering
    \includegraphics[width=\textwidth]{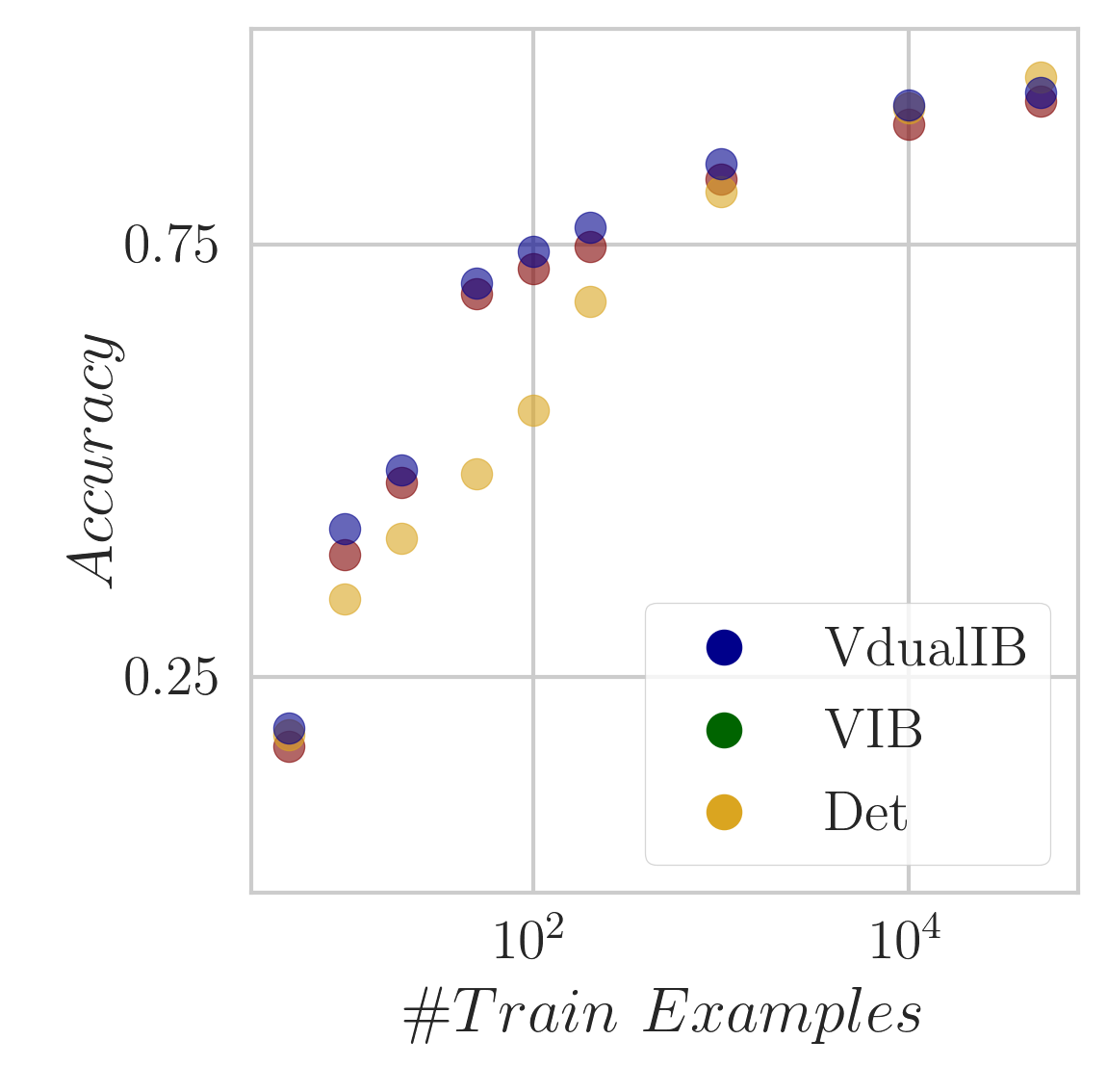}
    \caption{Accuracy vs. training size}
    \label{fig:train_examples_fasion}
\end{subfigure}
 \caption{Experiments over FashionMNIST. $(a)$ The information plane of the {\cvdib} and {\vib} for a range of $\beta$ values at the final training step. $(b)$ The evolution of the the {\cvdib}'s $I(X;\hX)$ along the optimization update steps. $(c)$ The models accuracy as a function of the training set size.}
 \label{fig:inf_plane_mnist}
\end{figure}

In \fref{fig:inf_plane_mnist} we present the information plane of the {\vdib} where the distribution $\tilde{p}\brk*{y \mid x}$ (the noise model) is a learnt confusion matrix, {\cvdib} (similarly to \citep{wu2020phase}). We compare it to the {\vib} over a range of $\beta$ values ($-5\leq\log{\beta} \leq 5$).
\fref{fig:inf_plane_mnist_beta} validates that, as expected, the information growth is approximately monotonic with $\beta$. Comparing the {\vdib} to the {\vib} model, we can see significant differences between their representations. The {\vdib} successfully obtains better compressed representations in comparison to the {\vib} performance, where only for large values of $I(X;\hX)$ their performances 
match.
As predicted by the theory, in the limit $\beta \rightarrow \infty$ the models behaviour match.
Furthermore, the {\vdib} values are smoother and they are spread over the information plane, making it easier to optimize for a specific value in it. In \fref{fig:inf_plane_mnist_compare} we consider the dynamics of $I(X;\hX)$  for several values of $\beta$. Interestingly, at the initial training stage the representation information for all values of $\beta$ decreases. However, as the training continues, the information increases only for high $\beta$s.

\subsubsection{The {\vdib} noise model}
\label{sec:cifar10_inf}
As mentioned above, learning with the {\vdib} objective requires 
 a choice of a noise model for the distribution $\tilde{p}\brk*{y\mid \hx}$. To explore the influence of different models on the learning we evaluate four types, with different assumptions on the access to the data. (i) Adding Gaussian noise to the one-hot vector of the true label ({\gvdib}); (ii) An analytic Gaussian integration of the log-loss around the one-hot labels; (iii) 
A pre-computed confusion matrix for the labels ({\cvdib}) as in  \citep{wu2020phase}; (iv) 
Using predictions of another trained model as the induced distribution. 
Where for (i) and (ii) the variance acts as a free parameter determining the noise level. The complexity of the noise models can be characterized by the additional prior knowledge on our data-set they require. While adding Gaussian noise does not require prior knowledge, using a trained model requires access to the prediction 
for every data sample. 
The use of a confusion matrix is an intermediate level 
of prior knowledge requiring access only to the $\abs{\mathcal{Y}}\times\abs{\mathcal{Y}}$ pre-computed matrix.
Here we present 
cases (i) and (iii) 
(see \S \ref{app:noise_model} for  (ii) and (iv)). Note that although using the {\vib} does not require the use of a noise model we incorporate it by replacing the labels with $\tilde{p}\brk*{y \mid x}$.
In the analysis below, the results are presented with the {\vib} trained with the same noise model as the {\vdib} (see \S \ref{app:noise_model} for  
a comparison between training {\vdib} with noise and {\vib} without it).

\begin{figure}
\begin{subfigure}{.3\textwidth}
    \centering
    \includegraphics[width=\textwidth]{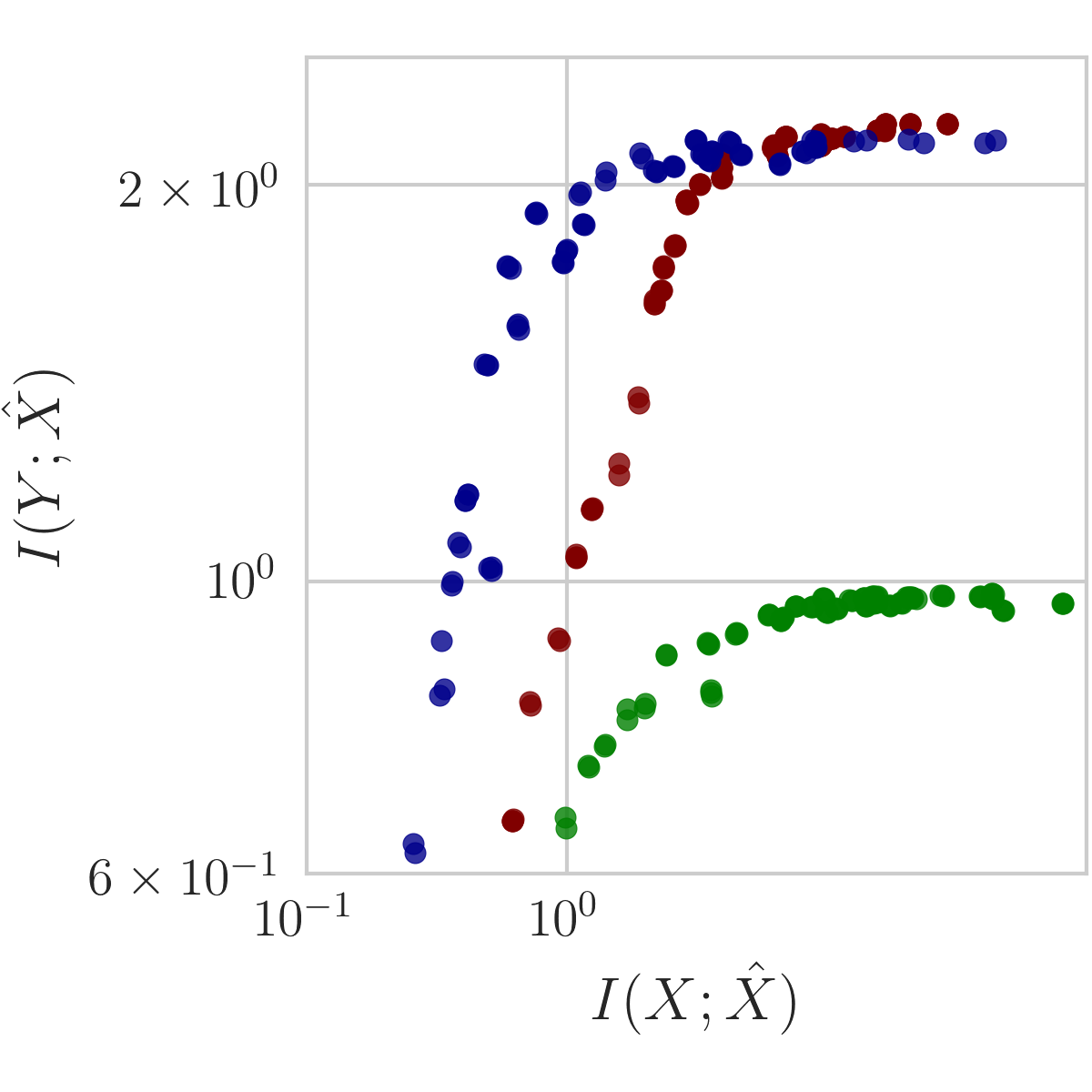}
    \caption{The information plane}
    \label{fig:inf_plane_all}
\end{subfigure}
\begin{subfigure}{.3\textwidth}
    \centering
    \includegraphics[width=\textwidth]{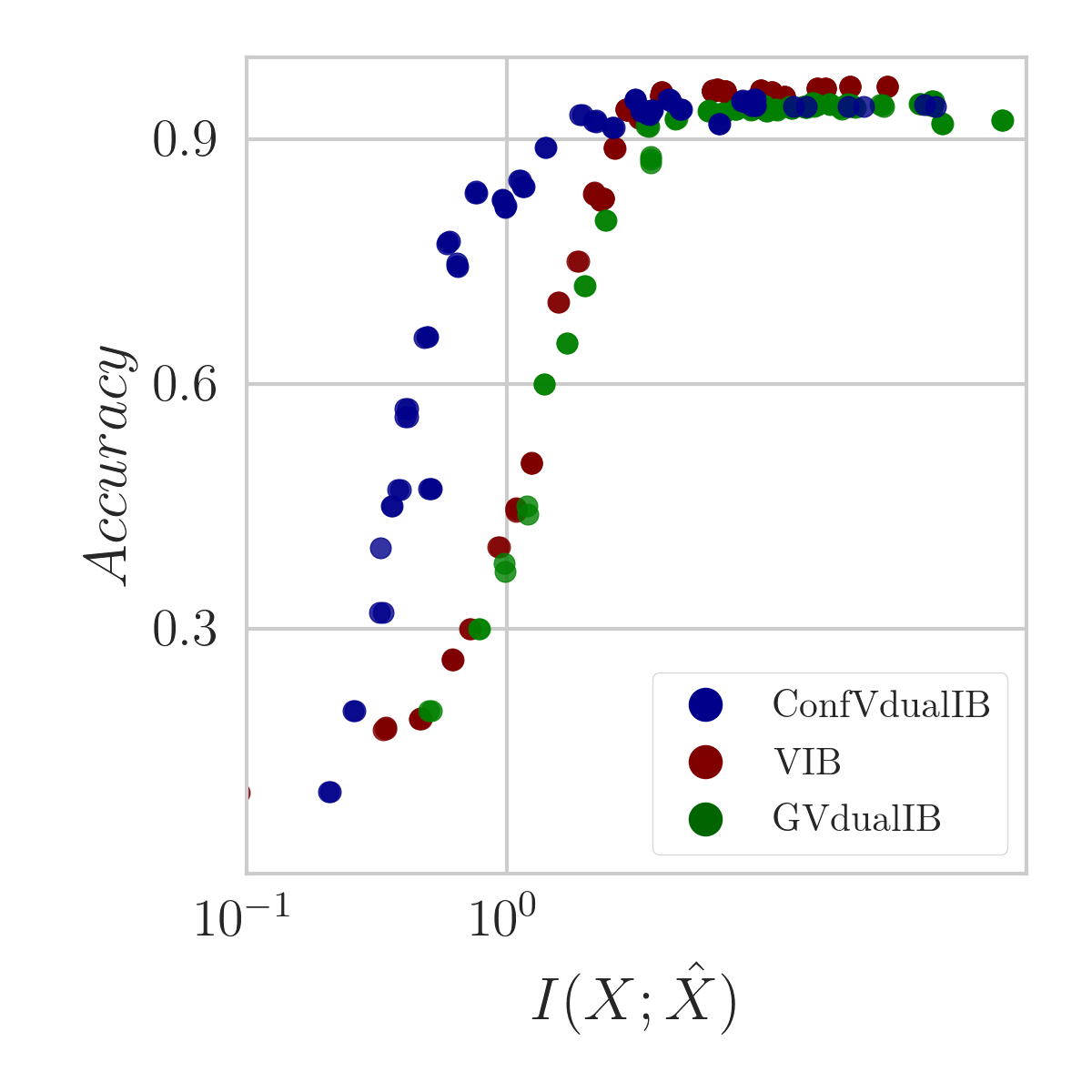}
    \caption{Accuracy vs. $I(X;\hX)$}
    \label{fig:inf_plane_acc_1}
\end{subfigure}
\begin{subfigure}{.3\textwidth}
    \centering
    \includegraphics[width=\textwidth]{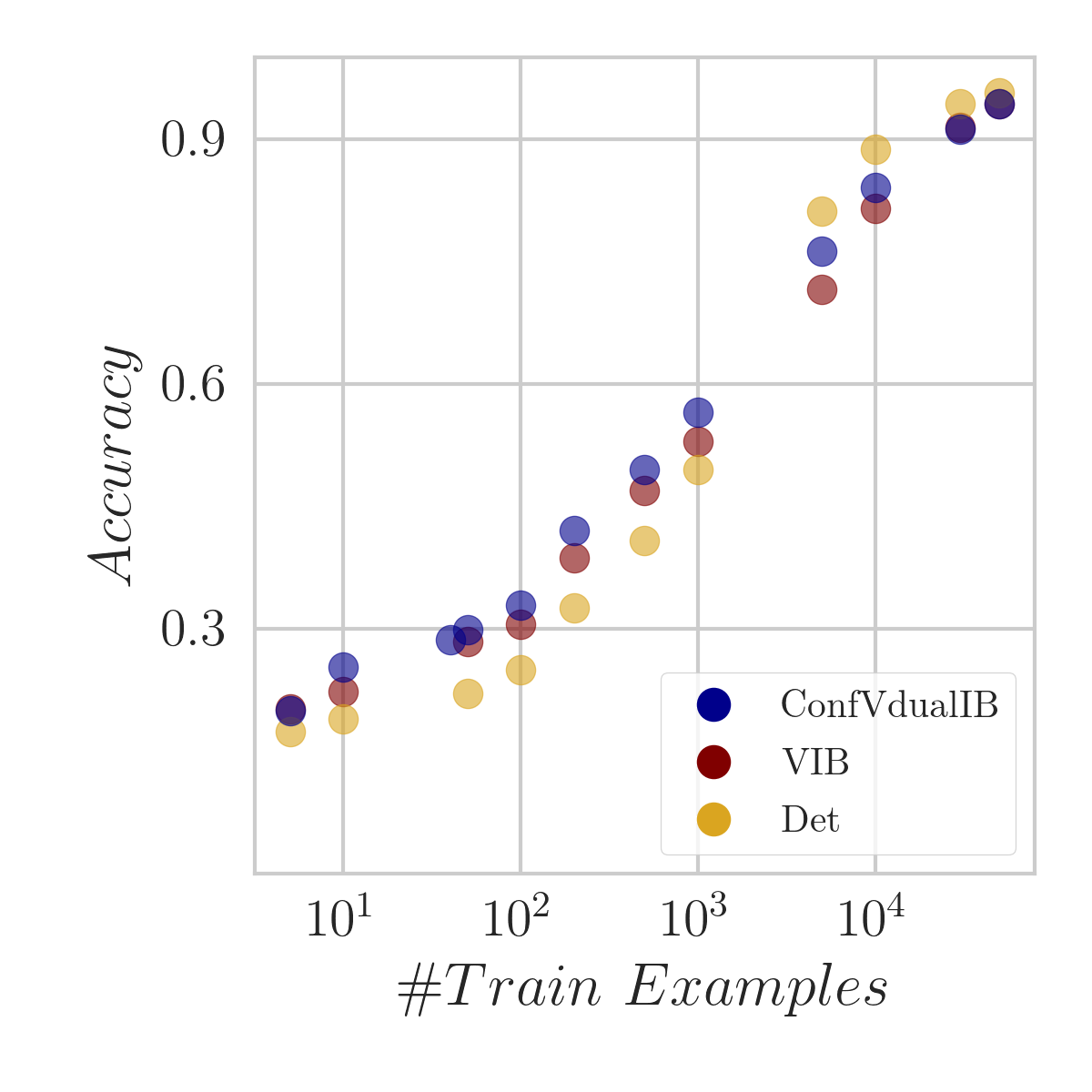}
    \caption{Accuracy vs. training size}
    \label{fig:train_examples_cifar}
\end{subfigure}
\caption{Experiments over CIFAR10. $(a)$ The information plane of the {\vib}, {\cvdib}, {\gvdib}  and {\vib} for a range of $\beta$ values. $(b)$ The accuracy of the models as a function of the mutual information, $I(X;\hX)$. $(c)$ The accuracy of the models as a function of the training set size.}
\label{fig:inf_plane}
\end{figure}

\fref{fig:inf_plane} depicts 
the information plane of the CIFAR10 data-set. \fref{fig:inf_plane_all} shows the information obtained from a range of $\beta$. The colors depict the different models {\cvdib}, {\gvdib} and the {\vib}. As we can see, training a {\vdib} with Gaussian noise achieves much less information with the labels at any given $I(X;\hX)$. We note that we verified that this behaviour is consistent over a wide range of variances.
The {\cvdib} model performance is similar to the {\vib}'s with the former showcasing more compressed representations. When we present the prediction accuracy (\fref{fig:inf_plane_acc_1}), here all  models attain roughly the same accuracy values. The discrepancy between the accuracy and information, $I(Y;\hX)$,
is similar to the one discussed in \citep{dusenberry2020efficient}. 

\subsubsection{Performance with different training set sizes}
\label{sec:experiment_setsize}
Our theoretical analysis (\S \ref{sec:min_err_exp}) shows that under given assumptions the {\dualib} bounds the optimal achievable error exponent on expectation hence it optimizes the error for a given data size $n$. We turn to test this in the {\vdib} setting. 
We train the models on a subset of the training set and evaluate them on the test set. We compare the {\vib} and the {\vdib} to a deterministic network (Det; Wide Res Net 28-10). Both the {\vib} and {\vdib} are trained over a wide range of $\beta$ values ($-5 \leq \log{\beta}\leq 6$). Presented is the best accuracy value for each model at a given $n$. 
\fref{fig:train_examples_fasion} and \fref{fig:train_examples_cifar} 
 show the accuracy of the models as a function of the training set size over FashionMNIST and CIFAR10 respectively. The {\vdib} performance is slightly better in comparison to the {\vib}, while the accuracy of the deterministic network is lower for small training sets. The superiority of the variational models over the deterministic network is not surprising as minimizing $I(X;\hX)$ acts as regularization.

\section{Conclusions}
\label{sec:conclusions}
We present here the Dual Information Bottleneck ({\dualib}), a framework resolving some of the known drawbacks of the {\ib} obtained by a mere switch between the terms in the distortion function. 
We provide the {\dualib} self-consistent equations allowing 
us to obtain analytical solutions. 
A local stability analysis revealed the underlying structure of the critical points of the solutions, resulting in a full bifurcation diagram of the optimal pattern representations.
The study of the {\dualib} objective reveals several interesting properties.  First, when the data can be modeled in a parametric form the {\dualib} preserves this structure and it obtains the representation in terms of the original parameters, as given by the {\expib} equations. Second, it optimizes the mean prediction error exponent thus improving the accuracy of the predictions as a function of the data size.
In addition to the {\dualib} analytic solutions, we provide a variational {\dualib} ({\vdib}) framework, which optimizes the functional using DNNs. This framework enables practical implementation of the {\dualib} to real world data-sets. While a broader analysis is required, the {\vdib} experiments shown validate the theoretical predictions. Our results demonstrate the potential advantages and unique properties of the framework.


\clearpage

   \setcounter{figure}{0}
        \renewcommand{\thefigure}{A\arabic{figure}}%
\appendix

\section*{Appendix}
\section{The Information Bottleneck method}
\label{app:ib}
The Information Bottleneck ({\ib}) trade off between the encoder and decoder mutual information values is defined by the minimization of the Lagrangian:
\begin{align}
\label{eq:IB_L1}
    \mathcal{F}\brk[s]*{p_{\beta}\brk*{\hat{x} \mid x}; p_{\beta}\brk*{y \mid \hat{x} }}  =I(X;\hat{X}) - \beta I(Y; \hat{X})~,
\end{align}

independently over the convex sets of the normalized distributions, $\brk[c]*{p_{\beta}\brk*{\hat{x} \mid x}}$, $\brk[c]*{p_{\beta}\brk*{\hat{x}}}$ and $\brk[c]*{p_{\beta}\brk*{y \mid \hat{x}}}$, given a positive Lagrange multiplier $\beta$. As shown in \cite{tishby99information, DBLP:conf/alt/ShamirST08}, this is a natural generalization of the classical concept of \emph{Minimal Sufficient Statistics} \cite{Cover:2006:EIT:1146355}, where the estimated parameter is replaced by the output variable $Y$ and \emph{exact} statistical sufficiency is characterized by the mutual information equality: $I(\hX;Y)=I(X;Y)$. The minimality of the statistics is captured by the minimization of $I(X;\hX) $, due to the Data Processing Inequality (DPI). However, non-trivial minimal sufficient statistics only exist for very special parametric distributions known as exponential families \cite{Exp_forms}. Thus in general, the {\ib} relaxes the minimal sufficiency problem to a continuous family of representations $\hX$ which are characterized by the trade off between compression, $I(X;\hX)\equiv I_{X}$, and accuracy, $I(Y;\hX)\equiv I_{Y}$, along a convex line in the \emph{Information-Plane} ($I_{Y}$ vs. $I_{X}$). When the rule $p(x,y)$ is strictly stochastic, the convex optimal line is smooth and each point along the line is uniquely characterized by the value of $\beta$. We can then consider the optimal representations $\hx=\hx(\beta)$ as encoder-decoder pairs: $(p_{\beta}(x\mid \hx),p_{\beta}(y\mid \hx))$\footnote{Here we use the \emph{inverse encoder}, which is in the fixed dimension simplex of distributions over $X$.} - a point in the continuous manifold defined by the Cartesian product of these distribution simplexes. We also consider a small variation of these representations, $\delta\hx$, as an infinitesimal change in this (encoder-decoder) continuous manifold (not necessarily on the optimal line(s)).         

\subsection{IB and Rate-Distortion Theory}

The {\ib} optimization trade off can be considered as a generalized rate-distortion problem \cite{Cover:2006:EIT:1146355} with the distortion function between a data point, $x$ and a representation point $\hx$ taken as the KL-divergence between their predictions of the desired label $y$:
\begin{align}
d_{\ib}\brk*{x,\hx}&=D\brk[s]*{p\brk*{y \mid x}||p_{\beta}\brk*{y|\hx}} \nonumber \\
&=\sum_{y}p\brk*{y \mid x}\log \frac{p\brk*{y \mid x}}{p_{\beta}\brk*{y \mid \hx}}.  
\end{align}
The expected distortion $\mathbb{E}_{p_{\beta}(x,\hx)}\brk[s]*{d_{\ib}\brk*{x,\hx}}$ for the optimal decoder is simply the label-information loss: $I(X;Y)-I(\hX;Y)$, using the Markov chain condition. Thus minimizing the expected $\ib$ distortion is equivalent to maximizing $I(\hX;Y)$, or minimizing \eqref{eq:IB_L}.  
Minimizing this distortion is equivalent to minimizing the cross-entropy loss, and it provides an upper-bound to other loss functions such as the $\mathcal{L}_{1}$-loss (due to the Pinsker inequality, see also \cite{Painsky2019}). Pinsker implies that both orders of the cross-entropy act as an upper bound to the $\mathcal{L}_{1}$-loss,
 $\min\{D\brk[s]*{q||p},D\brk[s]*{p||q}\} \geq \frac{1}{2 \log 2} \|p - q \|_{1}^{2}~.$  
 
\subsection{The IB Equations}

For discrete $X$ and $Y$, a necessary condition for the $\ib$ (local) minimization is given by the three self-consistent equations for the optimal encoder-decoder pairs, known as the \emph{{\ib} equations}:
\begin{align} \label{eq:IB}
	\begin{cases}
	\brk*{i}\ &p_{\beta}\brk*{\hat{x} \mid x} = \frac{p_{\beta}\brk*{\hat{x}}}{Z\brk*{x;\beta}} e^{-\beta D\brk[s]*{p\brk*{y\mid x} \| p_{\beta}\brk*{y \mid \hat{x}}}} \\
	\brk*{ii}\ &p_{\beta}\brk*{\hat{x}} = \sum_{x} p_{\beta}\brk*{\hat{x} \mid x} p\brk*{x} \\
	\brk*{iii}\ &p_{\beta}\brk*{y \mid \hat{x} } = \sum_{x} p\brk*{y\mid x} p_{\beta}\brk*{x \mid \hat{x} } 
	\end{cases}
,\end{align}
where $Z\brk*{x;\beta}$ is the normalization function. Iterating these equations is a generalized, Blahut-Arimoto, alternating projection algorithm \cite{CIS-58533,Cover:2006:EIT:1146355} and it converges to a stationary point of the Lagrangian, \eqref{eq:IB_L} \cite{tishby99information}. Notice that the minimizing decoder, (\eqref{eq:IB}-$(iii)$), is precisely the \emph{Bayes optimal decoder} for the representation $\hx(\beta)$, given the Markov chain conditions.

\subsection{Critical points and critical slowing down}
\label{sec:bifurcation_points}

One of the most interesting aspects of the {\ib} equations is the existence of critical points along the optimal line of solutions in the information plane (i.e. the information curve). At these points the representations change topology and cardinality (number of clusters) \cite{ZaslavskyTishby:2019, parker} and they form the skeleton of the information curve and representation space.  
To identify such points we perform a perturbation analysis of the {\ib} equations:\footnote{We ignore here the possible interaction between the different representations, for simplicity.}:
 \begin{align}
    \delta \log p_{\beta}\brk*{x \mid \hx}  =&  \beta\sum_{y} p\brk*{y\mid x} \delta \log p_{\beta}\brk*{y \mid \hx}
    \label{eq:stab_anl_x_ib} 
    ,\end{align}
    \begin{align}
    \delta \log p_{\beta}\brk*{y \mid \hx} = \frac{1}{p_{\beta}\brk*{y\mid \hx}} \sum_{x} p\brk*{y\mid x}p_{\beta}\brk*{x\mid \hx} \delta \log p_{\beta}\brk*{x \mid \hx}  \label{eq:stab_anl_y_ib}
.\end{align}
Substituting \eqref{eq:stab_anl_y_ib} into \eqref{eq:stab_anl_x_ib} and vice versa one obtains:
\begin{align*}
    \delta \log p_{\beta}\brk*{x \mid \hx}  &= \beta\sum_{y,x'}  p\brk*{y\mid x'} \frac{p\brk*{y\mid x} }{p_{\beta}\brk*{y\mid \hx}} p_{\beta}\brk*{x'\mid \hx} \delta \log p_{\beta}\brk*{x' \mid \hx} \\
    \delta \log p_{\beta}\brk*{y \mid \hx} &= \beta \sum_{x,y'} p\brk*{y\mid x} \frac{p_{\beta}\brk*{x\mid \hx}}{p_{\beta}\brk*{y\mid \hx}}   p\brk*{y'\mid x} \delta \log p_{\beta}\brk*{y' \mid \hx}
\end{align*}
Thus by defining the matrices: 
\begin{align} \label{eq:c_mat}
	C^{\ib}_{xx'}(\hx,\beta) &= \sum_{y}p\brk*{y \mid x} \frac{p_{\beta}\brk*{x' \mid \hat{x}} }{p_{\beta}\brk*{y \mid \hat{x} }} p\brk*{y \mid x' } ~,~ 
	C^{\ib}_{yy'}(\hx,\beta) = \sum_{x}p\brk*{y \mid x} \frac{p_{\beta}\brk*{x \mid \hat{x} }}{p_{\beta}\brk*{y \mid \hat{x} }} p\brk*{ y' \mid x }
.\end{align}
We obtain the following nonlinear eigenvalues problem:
\begin{align}
    \label{eq:stab_anl}
	\brk[s]*{I - \beta 	C^{\ib}_{xx'} \brk*{\hat{x}, \beta }} \delta \log p_{\beta}\brk*{x' \mid \hat{x}}&=0 ~,~~~~ 
	\brk[s]*{I - \beta C^{\ib}_{y y'}\brk*{\hat{x}, \beta }} \delta \log p_{\beta}\brk*{y' \mid \hat{x}}=0
,\end{align}

These two matrices have the same eigenvalues and have non-trivial eigenvectors  (i.e., different co-existing optimal representations) at the critical values of $\beta$, the bifurcation points of the {\ib} solution. At these points the cardinality of the representation $\hX$ (the number of ``{\ib}-clusters") changes due to splits of clusters, resulting in topological phase transitions in the encoder. These critical points form the ``skeleton" of the topology of the optimal representations. Between critical points the optimal representations change continuously (with $\beta$). The important computational consequence of critical points is known as \emph{critical slowing down} \cite{CriticalSlowingDown:2004}. 
For binary $y$, near a critical point the convergence time, $\tau_{\beta}$, of the iterations of \eqref{eq:IB} scales like:
$ \tau_{\beta} \sim {1}/{(1-\beta \lambda_2)}$,
where $\lambda_2$ is the second eigenvalue of either $C^{\ib}_{yy'}$ or $C^{\ib}_{xx'}$. At criticality, $\lambda_2(\hx)=\beta^{-1}$ and the number of iterations diverges. This phenomenon dominates any local minimization of \eqref{eq:IB} which is based on alternate encoder-decoder optimization. 

The appearance of the critical points and the critical slowing-down is visualized in Figure $1$ in the main text.

\section{The dualIB mathematical formulation}
\label{app:dual_ib}
The {\dualib} is solved with respect to the full Markov chain ($Y\rightarrow X \rightarrow \hX_{\beta} \rightarrow \hY$) in which we introduce the new variable, $\hy$, the \emph{predicted label}. Thus, in analogy to the $\ib$ we want to write the optimization problem in term of $\hY$.

Developing the expected distortion we find:
\begin{align*}
    \mathbb{E}_{p_{\beta}\brk*{x, \hat{x}}}\brk[s]*{d_{\dualib}\brk*{x, \hat{x}}} &= \sum_{x, \hat{x}}p_{\beta}\brk*{x, \hat{x}} \sum_{\hy} p_{\beta}\brk*{y=\hy \mid \hat{x}} \log \frac{p_{\beta}\brk*{y=\hy \mid \hat{x}} }{p\brk*{y=\hy \mid {x}}} \nonumber \\
     &=  \sum_{\hx, \hy} p_{\beta}\brk*{\hx} p_{\beta}\brk*{\hy \mid {\hx}} \log \frac{{p_{\beta}\brk*{\hy \mid \hx} }}{p_{\beta}\brk*{\hy}} - \sum_{x, \hy} p\brk*{x} p_{\beta}\brk*{\hy \mid {x}} \log \frac{{p_{\beta}\brk*{\hy \mid {x}} }}{p_{\beta}\brk*{\hy}} \nonumber \\
     &+ \sum_{x, \hy} p\brk*{x} p_{\beta}\brk*{\hy \mid {x}} \log \frac{p_{\beta}\brk*{\hy \mid {x}} }{p\brk*{y=\hy \mid {x}}} \nonumber \\
     &= I(\hX ; \hY) - I(X; \hY) + \mathbb{E}_{p\brk*{x }}\brk[s]*{D\brk[s]*{p_{\beta}\brk*{\hy \mid x} \| p\brk*{y=\hy \mid x} }}
.\end{align*}

Allowing the dual optimization problem to be written as:
  \begin{align*} 
      \mathcal{F}^{*}\brk[s]*{p\brk*{\hx \mid x} ;p\brk*{y \mid \hx }} &=  I(X;\hX) -\beta \brk[c]*{ I(X; \hY) - I(\hX ; \hY)- \mathbb{E}_{p\brk*{x }}\brk[s]*{D\brk[s]*{p_{\beta}\brk*{\hy \mid x} \| p\brk*{y=\hy \mid x} }}}
  .\end{align*}

\section{The DualIB solutions}
\label{app:dual_ib_eq_proof}
To prove \emph{theorem} $2$ we want to obtain the normalized distributions minimizing the {\dualib} rate-distortion problem.
\begin{proof} 
$(i)$ Given that the problem is formulated as a rate-distortion problem the encoder's update rule must be the known minimizer of the distortion function. \cite{Cover:2006:EIT:1146355}. 
Thus the $\ib$ encoder with the dual distortion is plugged in.
$(ii)$ For the decoder, by considering a small perturbation in the distortion $d_{\dualib}\brk*{x, \hat{x}}$, with $\alpha\brk*{\hat{x}}$ the normalization Lagrange multiplier, we obtain:
\begin{align*}
    \delta d_{\dualib}\brk*{x, \hat{x}} &= \delta \brk*{\sum_{y} p_{\beta}\brk*{y \mid \hat{x}} \log \frac{p_{\beta}\brk*{y \mid \hat{x}} }{p\brk*{y \mid {x}}} + \alpha\brk*{\hat{x}} \brk*{\sum_{y}p_{\beta}\brk*{y \mid \hat{x}} - 1 } }\nonumber \\
    \frac{\delta d_{\dualib}\brk*{x, \hat{x}} }{\delta p_{\beta}\brk*{y \mid \hat{x}}} &= \log \frac{p_{\beta}\brk*{y \mid \hat{x}} }{p\brk*{y \mid {x}}} + 1  + \alpha\brk*{\hat{x}} 
.\end{align*}
Hence, minimizing the expected distortion becomes:
\begin{align*}
    0 &= \sum_{x}  p_{\beta}\brk*{x \mid \hat{x}}\brk[s]*{\log \frac{p_{\beta}\brk*{y \mid \hat{x}} }{p\brk*{y \mid {x}}} + 1}  + \alpha\brk*{\hat{x}} \nonumber \\
    &= \log p_{\beta}\brk*{y \mid \hat{x}} - \sum_{x}  p_{\beta}\brk*{x \mid \hat{x}} \log p\brk*{y \mid x} + 1 + \alpha\brk*{\hat{x}} 
,\end{align*}
which yields Algorithm $1$, row $6$.
\end{proof}

Considering the $\dualib$ encoder-decoder, Algorithm $1$, we find that $\mathbb{E}_{p_{\beta}\brk*{x, \hat{x}}}\brk[s]*{d_{\dualib}\brk*{x, \hat{x}}}$ reduces to the expectation of the decoder's $\log$ partition function:
   \begin{align*} 
      \mathbb{E}_{p_{\beta}\brk*{x, \hat{x}}}\brk[s]*{d_{\dualib}\brk*{x, \hat{x}}} &= \sum_{x, \hat{x}}p_{\beta}\brk*{x, \hat{x}} \sum_{y} p_{\beta}\brk*{{y} \mid \hat{x}} \log \frac{p_{\beta}\brk*{{y} \mid \hat{x}} }{p\brk*{{y} \mid {x}}} \nonumber \\
      &= - \mathbb{E}_{p_{\beta}\brk*{\hx}}\brk[s]*{\log Z_{\rvy\mid \hat{\rvx}}\brk*{\hat{x}; \beta}} + \sum_{\hat{x}, y} p_{\beta}\brk*{\hx } \brk[s]*{ \sum_{x'} p_{\beta}\brk*{x'\mid \hat{x}} \log p\brk*{y \mid x'}  - \sum_{x} p_{\beta}\brk*{x\mid \hat{x}} \log p\brk*{y \mid x}} \nonumber \\
      &= - \mathbb{E}_{p_{\beta}\brk*{\hx}}\brk[s]*{\log Z_{\rvy\mid \hat{\rvx}}\brk*{\hat{x}; \beta}}
  .\end{align*}

\section{Stability analysis}
\label{app:stability_anl}
Here we provide the detailed stability analysis allowing the definition of the matrices $C^{\dualib}_{xx'}, C^{\dualib}_{yy'}$ (\emph{theorem $4$}) which allows us to claim that they obey the same rules as the $C$ matrices of the {\ib}. Similarly to the  {\ib} in this calculation we ignore second order contributions which arise form the normalization terms.
Considering a variation in $\hat{x}$ we get:
 \begin{align}
    \delta \log p_{\beta}\brk*{x \mid \hx} =& \beta\sum_{y}p_{\beta}\brk*{y \mid \hat{x}}   \brk*{\log \frac{ p\brk*{y \mid x} }{p_{\beta} \brk*{y \mid \hat{x}}} - 1}\delta \log p_{\beta}\brk*{y \mid \hx}\nonumber \\
    =&  \beta\sum_{y}p_{\beta}\brk*{y \mid \hat{x}} \brk[s]*{\log  p\brk*{y \mid x} - \sum_{\tilde{x}} p_{\beta} \brk*{\tilde{x} \mid \hx}\log  p\brk*{y \mid \tilde{x}} }
    \delta \log p_{\beta}\brk*{y \mid \hx} \nonumber \\
    +& \beta\sum_{y} \log  Z_{\rvy\mid \hat{\rvx}}\brk*{\hat{x}; \beta} \frac{\partial  p_{\beta}\brk*{y \mid \hat{x}}}{\partial \hat{x}} \nonumber \\
    =&  \beta\sum_{y,\tilde{x}}p_{\beta}\brk*{y \mid \hat{x}} p_{\beta} \brk*{\tilde{x} \mid \hx} \log \frac{ p\brk*{y \mid x} }{p \brk*{y \mid \tilde{x}}}\delta \log p_{\beta}\brk*{y \mid \hx}
    \label{eq:stab_anl_x} 
    ,\end{align}
    \begin{align}
    \delta \log p_{\beta}\brk*{y \mid \hx} =&- \frac{1}{ Z_{\rvy\mid \hat{\rvx}}\brk*{\hat{x}; \beta}}\frac{\partial Z_{\rvy\mid \hat{\rvx}}\brk*{\hat{x}; \beta}}{\partial \hat{x}}+  \sum_{x}p_{\beta}\brk*{x \mid \hat{x}}\log p\brk*{y \mid x}\delta \log p_{\beta}\brk*{x \mid \hx} \nonumber \\
    =&  -\sum_{\tilde{y}} p_{\beta}\brk*{\tilde{y} \mid \hat{x}} \sum_{x}p_{\beta}\brk*{x \mid \hat{x}} \log p\brk*{\tilde{y} \mid x} \delta \log p_{\beta}\brk*{x \mid \hx} \nonumber \\
    +&\sum_{x} p_{\beta}\brk*{x \mid \hat{x}} \log p\brk*{y \mid x}\delta \log p_{\beta}\brk*{x \mid \hx} 
    \nonumber \\
    =& \sum_{x, \tilde{y}} p_{\beta}\brk*{x \mid \hat{x}}p_{\beta}\brk*{\tilde{y} \mid \hat{x}}\log \frac{{p\brk*{y \mid x}}}{p\brk*{\tilde{y} \mid x}} \delta \log p_{\beta}\brk*{x \mid \hx} \label{eq:stab_anl_y}
.\end{align}

Substituting \eqref{eq:stab_anl_y} into \eqref{eq:stab_anl_x} and vice versa one obtains:
 \begin{align*} 
    \delta \log p_{\beta}\brk*{x \mid \hx} &= \beta\sum_{x',y,\tilde{y}, \tilde{x}}p_{\beta}\brk*{y \mid \hat{x}} p_{\beta} \brk*{\tilde{x} \mid \hat{x}}  \log \frac{ p\brk*{y \mid x} }{p \brk*{y \mid \tilde{x}}}   \nonumber \\
    &\cdot p_{\beta}\brk*{x' \mid \hat{x}}p_{\beta}\brk*{\tilde{y} \mid \hat{x}}  \log \frac{p\brk*{y \mid x'}}{p\brk*{\tilde{y} \mid x'} } {\delta \log p_{\beta}\brk*{x' \mid \hat{x}}} \nonumber \\
    \delta \log p_{\beta}\brk*{y \mid \hx} &= \beta \sum_{x ,y', \tilde{x}, \tilde{y}} p_{\beta}\brk*{x \mid \hat{x}}p_{\beta}\brk*{\tilde{y} \mid \hat{x}}  \log \frac{p\brk*{y \mid x}}{p\brk*{\tilde{y} \mid x} }  \nonumber \\
    &\cdot p_{\beta}\brk*{y' \mid \hat{x}} p_{\beta}\brk*{\tilde{x} \mid \hat{x}} \log \frac{ p\brk*{y' \mid x} }{p \brk*{y' \mid \tilde{x}}} {\delta \log p_{\beta}\brk*{y' \mid \hat{x}}}
.\end{align*}

We now define the $C^{\dualib}$ matrices as follows:
\begin{align*} 
	C^{\dualib}_{xx'}\brk*{\hat{x}; \beta} = &\sum_{y, \tilde{y}, \tilde{x}} p_{\beta}\brk*{y \mid \hat{x}}  p_{\beta}\brk*{\tilde{x} \mid \hat{x}}   \log\frac{ p\brk*{y \mid x} }{p \brk*{y \mid \tilde{x}}}  \cdot  p_{\beta}\brk*{x' \mid \hat{x}}p_{\beta}\brk*{\tilde{y} \mid \hat{x}} \log \frac{p\brk*{y \mid x'}}{p\brk*{\tilde{y} \mid x'} }
	 \nonumber \\
	C^{\dualib}_{yy'}\brk*{\hat{x}; \beta} = &\sum_{x, \tilde{x}, \tilde{y}} p_{\beta}\brk*{x \mid \hat{x}}  p_{\beta}\brk*{\tilde{y} \mid \hat{x}} \log \frac{p\brk*{y \mid x}}{p\brk*{\tilde{y} \mid x}}  
	\cdot p_{\beta}\brk*{y' \mid \hat{x}} p_{\beta}\brk*{\tilde{x} \mid \hat{x}} \log \frac{ p\brk*{y' \mid x} }{p \brk*{y' \mid \tilde{x}}} 
.\end{align*}
Using the above definition we have an equivalence to the {\ib} stability analysis in the form of: 
\begin{align*} 
	\brk[s]*{I - \beta 	C^{\dualib}_{xx'} \brk*{\hat{x}, \beta }} {\delta \log p_{\beta}\brk*{x' \mid \hat{x}}}&=0 ~,~~~~ 
	\brk[s]*{I - \beta C^{\dualib}_{y y'}\brk*{\hat{x}, \beta }} {\delta \log p_{\beta}\brk*{y' \mid \hat{x}}}=0
.\end{align*}
Note that for the binary case, the matrices may be simplified to:
\begin{align*} 
	C^{\dualib}_{xx'}\brk*{\hat{x}; \beta} = &\sum_{y, \tilde{x}} p_{\beta}\brk*{y \mid \hat{x}}  p_{\beta}\brk*{\tilde{x} \mid \hat{x}}   \log\frac{ p\brk*{y \mid x} }{p \brk*{y \mid \tilde{x}}}  \cdot  p_{\beta}\brk*{x' \mid \hat{x}}\brk*{1-p_{\beta}\brk*{y \mid \hat{x}}} \log \frac{p\brk*{y \mid x'}}{1-p\brk*{y \mid x'} }
	 \nonumber \\
	C^{\dualib}_{yy'}\brk*{\hat{x}; \beta} = &\sum_{x, \tilde{x}} p_{\beta}\brk*{x \mid \hat{x}}  \brk*{1-p_{\beta}\brk*{y \mid \hat{x}}} \log \frac{p\brk*{y \mid x}}{1-p\brk*{y \mid x}}  
	\cdot p_{\beta}\brk*{y' \mid \hat{x}} p_{\beta}\brk*{\tilde{x} \mid \hat{x}} \log \frac{ p\brk*{y' \mid x} }{p \brk*{y' \mid \tilde{x}}} 
.\end{align*}

We turn to show that the $C^{\dualib}$ matrices share the same eigenvalues with $\lambda_{1}\brk*{\hx} = 0$. 
\begin{proof}
The matrices, $C^{\dualib}_{xx'}\brk*{\hat{x}; \beta}$, $C^{\dualib}_{yy'}\brk*{\hat{x}; \beta}$, are given by: 
\begin{equation*}
    C^{\dualib}_{xx'}\brk*{\hat{x}; \beta} = A_{xy}\brk*{\hx;\beta} B_{yx'}\brk*{\hx;\beta} ~,~  
	C^{\dualib}_{yy'}\brk*{\hat{x}; \beta} = B_{yx}\brk*{\hx;\beta} A_{xy'}\brk*{\hx;\beta}
, \end{equation*}
with: \\
\begin{align*} 
	A_{xy}\brk*{\hx;\beta} = p_{\beta}\brk*{y \mid \hat{x}}\sum_{\tilde{x}}p_{\beta}\brk*{\tilde{x} \mid \hat{x}}   \log\frac{ p\brk*{y \mid x} }{p \brk*{y \mid \tilde{x}}} ~, ~
	B_{yx}\brk*{\hx;\beta} = p_{\beta} \brk*{x \mid \hat{x}} \sum_{\tilde{y}}p_{\beta}\brk*{\tilde{y} \mid \hat{x}}\log \frac{{p\brk*{y \mid x}}}{ p\brk*{\tilde{y} \mid x}}
.\end{align*}
Given that the matrices are obtained by the multiplication of the same matrices, it follows that they have the same eigenvalues $\brk[c]*{\lambda_i\brk*{\hx;\beta}}$. 

To prove that $\lambda_1\brk*{\hx;\beta} = 0$ we show that $\det(C^{\dualib}_{y y'})= 0$. We present the exact calculation for a binary label $y \in \brk[c]*{y_{0},y_{1}}$ (the argument for general $y$ follows by encoding the label as a sequence of bits and discussing the first bit only, as a binary case):
\begin{align*}
    	\det(C^{\dualib}_{yy'}\brk*{\hat{x}; \beta}) =& \sum_{x, \tilde{x}} p_{\beta}\brk*{x \mid \hat{x}}  p_{\beta}\brk*{y_{1} \mid \hat{x}} \log \frac{p\brk*{y_{0} \mid x}}{p\brk*{y_{1} \mid x}}  
	\cdot p_{\beta}\brk*{y_{0} \mid \hat{x}} p_{\beta}\brk*{\tilde{x} \mid \hat{x}} \log \frac{ p\brk*{y_{0} \mid x} }{p \brk*{y_{0} \mid \tilde{x}}} \\
	\cdot& \sum_{x', \tilde{x}',} p_{\beta}\brk*{{x'} \mid \hat{x}}  p_{\beta}\brk*{y_{0} \mid \hat{x}} \log \frac{p\brk*{y_{1} \mid {x'}}}{p\brk*{y_{0} \mid {x'}}}  
	\cdot p_{\beta}\brk*{y_{1} \mid \hat{x}} p_{\beta}\brk*{\tilde{x}' \mid \hat{x}} \log \frac{ p\brk*{y_{1} \mid {x'}} }{p \brk*{y_{1} \mid \tilde{x}'}} \\
	-& \sum_{x, \tilde{x}} p_{\beta}\brk*{x \mid \hat{x}}  p_{\beta}\brk*{y_{0} \mid \hat{x}} \log \frac{p\brk*{y_{1} \mid x}}{p\brk*{y_{0} \mid x}}  
	\cdot p_{\beta}\brk*{y_{0} \mid \hat{x}} p_{\beta}\brk*{\tilde{x} \mid \hat{x}} \log \frac{ p\brk*{y_{0} \mid x} }{p \brk*{y_{0} \mid \tilde{x}}} \\
	\cdot& \sum_{{x'}, \tilde{x}'} p_{\beta}\brk*{{x'} \mid \hat{x}}  p_{\beta}\brk*{y_{1} \mid \hat{x}} \log \frac{p\brk*{y_{0} \mid {x'}}}{p\brk*{y_{1} \mid {x}'}}  
	\cdot p_{\beta}\brk*{y_{1} \mid \hat{x}} p_{\beta}\brk*{\tilde{x}' \mid \hat{x}} \log \frac{ p\brk*{y_{1} \mid {x'}} }{p \brk*{y_{1} \mid \tilde{x}'}} \\
	=& \sum_{x, x', \tilde{x}, \tilde{x}'}
	p_{\beta}\brk*{x \mid \hat{x}}  p_{\beta}\brk*{{x'} \mid \hat{x}}  p^2_{\beta}\brk*{y_{0} \mid \hat{x}}  p^2_{\beta}\brk*{y_{1} \mid \hat{x}}  p_{\beta}\brk*{\tilde{x} \mid \hat{x}}\log \frac{ p\brk*{y_{0} \mid x} }{p \brk*{y_{0} \mid \tilde{x}}}  p_{\beta}\brk*{\tilde{x}' \mid \hat{x}} \log \frac{ p\brk*{y_{1} \mid {x'}} }{p \brk*{y_{1} \mid \tilde{x}'}}\\
	\cdot&  \brk[s]*{\log \frac{p\brk*{y_{0} \mid x}}{p\brk*{y_{1} \mid x}} \log \frac{p\brk*{y_{1} \mid {x'}}}{p\brk*{y_{0} \mid {x'}}}   -   \log \frac{ p\brk*{y_{0} \mid x} }{p \brk*{y_{1} \mid x}} \log \frac{p\brk*{y_{1} \mid {x'}}}{p\brk*{y_{0} \mid {x'}}}   } = 0
.\end{align*}
Given that the determinant is $0$ implies that $\lambda_{1}\brk*{\hx} = 0$. 
\end{proof}
For a binary problem we can describe the non-zero eigenvalue using $\lambda_{2}\brk*{\hx} =  \textrm{Tr}(C^{\dualib}_{yy'}\brk*{\hat{x}; \beta})$. That is:
\begin{align*}
    \lambda_{2}\brk*{\hx} =& \sum_{x, \tilde{x}} p_{\beta}\brk*{x \mid \hat{x}}  p_{\beta}\brk*{y_{1} \mid \hat{x}} \log \frac{p\brk*{y_{0} \mid x}}{p\brk*{y_{1} \mid x}}  
	\cdot p_{\beta}\brk*{y_{0} \mid \hat{x}} p_{\beta}\brk*{\tilde{x} \mid \hat{x}} \log \frac{ p\brk*{y_{0} \mid x} }{p \brk*{y_{0} \mid \tilde{x}}} \\
	+& \sum_{x, \tilde{x}} p_{\beta}\brk*{{x} \mid \hat{x}}  p_{\beta}\brk*{y_{0} \mid \hat{x}} \log \frac{p\brk*{y_{1} \mid {x}}}{p\brk*{y_{0} \mid {x}}}  
	\cdot p_{\beta}\brk*{y_{1} \mid \hat{x}} p_{\beta}\brk*{\tilde{x} \mid \hat{x}} \log \frac{ p\brk*{y_{1} \mid {x}} }{p \brk*{y_{1} \mid \tilde{x}}} \\
	=& p_{\beta}\brk*{y_{1} \mid \hat{x}}  
	p_{\beta}\brk*{y_{0} \mid \hat{x}}  \sum_{x, \tilde{x}} p_{\beta}\brk*{x \mid \hat{x}}  p_{\beta}\brk*{\tilde{x} \mid \hat{x}} \log \frac{p\brk*{y_{0} \mid x}}{p\brk*{y_{1} \mid x}} \brk[s]*{\log \frac{ p\brk*{y_{0} \mid x} }{p \brk*{y_{0} \mid \tilde{x}}} - \log \frac{ p\brk*{y_{1} \mid x} }{p \brk*{y_{1} \mid \tilde{x}}} }
.\end{align*}

\subsection{Definition of the sample problem}
\label{app:def_prob}
We consider a problem for a binary label $Y$ and $5$ possible inputs $X$ uniformly distributed, i.e.  $\forall x \in \mathcal{X}, p\brk*{x} = 1/5$ and the conditional distribution, $p\brk*{y\mid x}$, given by:
\begin{center}
            \begin{tabular}{c| c c c c c}
        & $x= 0$ & $x=1$ & $x=2 $ & $x=3$ & $x=4$ \\ \hline 
            $y=0$ &  0.12 & 0.23 & 0.4 & 0.6 & 0.76 \\ \hline 
            $y=1$ &  0.88 & 0.77 & 0.6 & 0.4 & 0.24 
        \end{tabular}
\end{center}

\section{Information plane analysis}
\label{app:performance_anl}
We rely on known results for the rate-distortion problem and the information plane:
\begin{lemma} \label{lm:dist_convex}
 $I(x;\hX)$ is a non-increasing convex function of the distortion $\mathbb{E}_{p_{\beta}\brk*{x, \hx}}\brk[s]*{d\brk*{x, \hx}}$ with a slope of $-\beta$. 
\end{lemma} 
We emphasis that this is a general result of rate-distortion thus holds for the $\dualib$ as well.
\begin{lemma} \label{lm:ib_concave}
For a fixed encoder $p_{\beta}\brk*{\hx \mid x}$ and the Bayes optimal decoder $p_{\beta}\brk*{y \mid \hx}$:
\begin{align*}
    \mathbb{E}_{p_{\beta}\brk*{x, \hx}}\brk[s]*{d_{\ib}\brk*{x, \hx}} = I(X ; Y) - I(\hX ; Y)
.\end{align*}
Thus, the information curve, $I_{y}$ vs. $I_{x}$, is a non-decreasing concave function with a positive slope, $\beta^{-1}$. The concavity implies that $\beta$ increases along the curve.
\end{lemma} 
\cite{Cover:2006:EIT:1146355, Gilad-bachrach}.

\subsection{Proof of Lemma 3} 
\label{sec:lemma_info_proof}
In the following section we provide a proof to \emph{lemma} $3$,  for the $\ib$ and {\dualib} problems.
\begin{proof}
We want to analyze the behavior of $I_{x}\brk*{\beta}$, $I_{y}\brk*{\beta}$, that is the change in each term as a function of the corresponding $\beta$. From \lmref{lm:ib_concave}, the concavity of the information curve, we can deduce that both are non-decreasing functions of $\beta$. As the two $\beta$ derivatives are proportional it's enough to discuss the first one. 

Next, we focus on their behavior between two critical points. That is, where the cardinality of $\hX$ is fixed (clusters are "static"). 
For ''static" clusters, the $\beta$ derivative of $I_{x}$, along the optimal line is given by:
\begin{align*} 
 \frac{\partial  I(X ; \hat{X})  }{\partial \beta} &=-  \frac{\partial  }{\partial\beta} \brk[s]*{\sum_{x, \hat{x}} p_{\beta}\brk*{ x, \hat{x}} \brk*{\log {Z_{\hat{\rvx} \mid \rvx }\brk*{x;\beta} } + \beta d\brk*{x,\hx} } } \\
 &= - \beta \brk[a]*{ d\brk*{{x} ,\hx }\frac{\partial \log p_{\beta} \brk*{\hx \mid x}  }{\partial \beta}  }_{p_{\beta}\brk*{{x} ,\hx }} \\ 
 &\approx  \beta \brk[a]*{ d\brk*{{x} ,\hx } \brk[s]*{ \frac{\partial \log Z_{\hat{\rvx} \mid \rvx }\brk*{x;\beta}  }{\partial \beta} + d\brk*{{x} ,\hx } }  }_{p_{\beta}\brk*{{x} ,\hx }} \\
&\approx  \beta\brk[a]*{\underbrace{\brk[a]*{ d^2\brk*{{x} ,\hx }}_{p_{\beta}\brk*{\hx \mid x} } -\brk[a]*{ d\brk*{{x} ,\hx } }^2_{p_{\beta}\brk*{\hx \mid x } } }_{\textrm{Var}\brk*{d\brk*{x}} } }_{p\brk*{x}}
.\end{align*}
This first of all reassures that the function is non-decreasing as $\textrm{Var}\brk*{d(x)} \geq 0$.

The piece-wise concavity follows from the fact that when the number of clusters is fixed (between the critical points) - increasing $\beta$ decreases the clusters conditional entropy $H(\hX \mid x)$, as the encoder becomes more deterministic. The mutual information is bounded by $H(\hX)$ and it's $\beta$ derivative decreases. Further, between the critical points there are no sign changes in the second $\beta$ derivative.
\end{proof}

\subsection{Proof of Theorem 4} 
\begin{proof}
The proof follows from \textit{lemma} $3$ together with the critical points analysis above, and is only sketched here. 
As the encoder and decoder at the critical points, $\beta_{c}^{\ib}$ and $\beta_{c}^{\dualib}$, have different left and right derivatives, they form cusps in the curves of the mutual information ($I_{x}$ and $I_{y}$) as functions of $\beta$. These cusps can only be consistent with the optimality of the {\ib} curves ( implying that sub-optimal curves lie below it; i.e, the {\ib} slope is steeper) if $\beta_{c}^{\dualib} < \beta_{c}^{\ib}$ (this is true for any sub-optimal distortion), otherwise the curves intersect. 

Moreover, at the $\dualib$ critical points, the distance between the curves is minimized due to the strict concavity of the functions segments between the critical points. As the critical points imply discontinuity in the derivative, this results in a ''jump" in the information values. Therefore, at any $\beta_{c}^{\dualib}$ the distance between the curves has a (local) minimum.
This is depicted in Figure $4$ (in the main text), comparing $I_{x}\brk*{\beta}$ and $I_{y}\brk*{\beta}$ and their differences for the two algorithms.

The two curves approach each other for large $\beta$ since the two distortion functions become close in the low distortion limit (as long as $p\brk*{y \mid x}$ is bounded away from $0$).
\end{proof}

\section{Derivation of the dualExpIB}
\label{app:dualexpIB}
We  provide elaborate derivations to \textit{theorem} $9$; that is, we obtain the {\dualib} optimal encoder-decoder under the exponential assumption over the data.
We use the notations defined in \S \emph{The Exponential Family dualIB}.
\begin{itemize}
    \item The \textit{decoder}. 
    Substituting the exponential assumption into the {\dualib} $\log$-decoder yields:
    \begin{align*} 
	\log p_{\beta}\brk*{y \mid \hat{x} } &=  \sum_{x} p_{\beta}\brk*{x \mid \hat{x}} \log p\brk*{y \mid x}  - \log Z_{\rvy\mid \hat{\rvx}} \brk*{\hat{x}; \beta}  \nonumber \\
	&= - \sum_{x} \sum_{r=0}^{d} p_{\beta}\brk*{x \mid \hat{x}}   {\lambda}^r\brk*{y} A_r \brk*{x} - \log Z_{\rvy\mid \hat{\rvx}} \brk*{\hat{x};\beta} \nonumber \\
	&= -  \sum_{r=1}^{d}  {\lambda}^r\brk*{y} A_{r,\beta} \brk*{\hx} - \mathbb{E}_{p_{\beta}\brk*{x\mid \hat{x}}} \brk[s]*{ {\lambda}^0_{\rvx}}- \log Z_{\rvy\mid \hat{\rvx}} \brk*{\hat{x};\beta}
.\end{align*}
Taking a closer look at the normalization term:
\begin{align*} 
	  Z_{\rvy\mid \hat{\rvx}} \brk*{\hat{x}; \beta} &= \sum_{y} e^{\sum_{x} p_{\beta}\brk*{x \mid \hat{x}} \log p\brk*{y \mid x}   } = e^{-\mathbb{E}_{p_{\beta}\brk*{x\mid \hat{x}}} \brk[s]*{ {\lambda}^0_{\rvx}}}\sum_{y} e^{-\sum_{r=1}^{d}  {\lambda}^r\brk*{y} A_{r, \beta} \brk*{\hx}} \nonumber \\
	  \log Z_{\rvy\mid \hat{\rvx}} \brk*{\hat{x};\beta} &= -\mathbb{E}_{p_{\beta}\brk*{x\mid \hat{x}}} \brk[s]*{ {\lambda}^0_{\rvx}} + \log \brk*{\sum_{y}e^{-\sum_{r=1}^{d}  {\lambda}^r\brk*{y} A_{r, \beta} \brk*{\hx}}} \nonumber
.\end{align*}
From which it follows that ${\lambda}^{0}_{\beta}\brk{ \hx}$ is given by:
\begin{align*} 
 {\lambda}^{0}_{\beta}\brk{\hx} &= \log \brk*{\sum_{y} e^{-\sum_{r=1}^{d}  {\lambda}^r\brk*{y} A_{r,\beta} \brk*{\hat{x}}}}
,\end{align*}
and we can conclude that the $\expib$ decoder takes the form:
\begin{align*} 
	\log p_{\beta}\brk*{y \mid \hat{x} } &=  -\sum_{r=1}^{d}  {\lambda}^r\brk*{y} A_{r,\beta}\brk*{\hat{x}} - {\lambda}^{0}_{\beta}\brk{\hx}
.\end{align*}
\item The \textit{encoder}. \\
The core of the encoder is the dual distortion function which may now be written as:
 \begin{align*} 
	 d_{\dualib}\brk*{x, \hat{x}} &= \sum_{y} p_{\beta}\brk*{y\mid \hat{x}} \log \frac{p_{\beta}\brk*{y\mid \hat{x}}}{{p}\brk*{y\mid {x}} } \nonumber \\
	 &=  \sum_{y} p_{\beta}\brk*{y\mid \hat{x}}\brk[s]*{ \brk*{{\lambda}^{0}_{\rvx}-{\lambda}^{0}_{\beta}\brk*{\hat{x}}}+\sum_{r=1}^{d} {\lambda}^r\brk*{y} \brk*{A_r \brk*{x} - A_{r,\beta}\brk*{\hat{x}} }}\nonumber\\
	 &= {\lambda}^{0}_{\rvx}-{\lambda}^{0}_{\beta}\brk*{\hx}  +  \sum_{r=1}^{d} {\lambda}_{\beta}^r\brk*{\hat{x}}   \brk*{A_r \brk*{x} - A_{r, \beta}\brk*{\hat{x}}} 
,\end{align*}
substituting this into the encoder's definition we obtain:
 \begin{align*} 
	 p_{\beta}\brk*{\hx\mid x} &= \frac{p_{\beta}\brk*{\hx}}{Z_{\hat{\rvx}\mid \rvx}\brk*{x;\beta} }  e^{ -\beta\brk[s]*{ {\lambda}^{0}_{\rvx}-{\lambda}^{0}_{\beta}\brk*{\hat{x}}+  \sum_{r=1}^{d} {\lambda}_{\beta}^r\brk*{\hat{x}}   \brk[s]*{A_r \brk*{x} - A_{r, \beta}\brk*{\hat{x}}} }} \\
	 &= \frac{p_{\beta}\brk*{\hx}e^{\beta {\lambda}^{0}_{\beta}\brk*{\hx}}}{Z_{\hat{\rvx}\mid \rvx}\brk*{x;\beta} }  e^{-\beta  \sum_{r=1}^{d} {\lambda}_{\beta}^r\brk*{\hat{x}}   \brk[s]*{A_r \brk*{x} - A_{r, \beta}\brk*{\hat{x}}} }
.\end{align*}
\end{itemize}
We can further write down the information quantities under these assumptions:
\begin{align*}
 I(X;\hX) &= \sum_{x,\hx} p_{\beta}\brk*{x, \hx } \log \frac{p_{\beta}\brk*{x \mid \hx}}{p\brk*{x } }  \\
        &= H\brk*{X} - \beta \sum_{r=1}^{d}  \sum_{\hx} p_{\beta}\brk*{\hx } {\lambda}_{\beta}^r\brk*{\hat{x}}   \brk[s]*{  \sum_{x} p_{\beta}\brk*{x\mid \hx } A_r \brk*{x} - A_{r, \beta}\brk*{\hat{x}}} + \beta  \mathbb{E}_{p_{\beta}\brk*{\hx}}\brk[s]*{{\lambda}^{0}_{\beta}\brk*{\hx} }  - \mathbb{E}_{p\brk*{x}}\brk[s]*{ \log Z_{\hat{\rvx}\mid \rvx}\brk*{x;\beta}} \\
        &= H\brk*{X} + \beta  \mathbb{E}_{p_{\beta}\brk*{\hx}}\brk[s]*{{\lambda}^{0}_{\beta}\brk*{\hx} }  - \mathbb{E}_{p\brk*{x}}\brk[s]*{ \log Z_{\hat{\rvx}\mid \rvx}\brk*{x;\beta}} \\
 I(Y;\hX) &= \sum_{y,\hx} p_{\beta}\brk*{y, \hx } \log \frac{p_{\beta}\brk*{y \mid \hx}}{p\brk*{y } }  \\
        &= H\brk*{Y} - \sum_{r=1}^{d} \sum_{\hx} p_{\beta}\brk*{ \hx }   \sum_{y}p_{\beta}\brk*{y \mid \hx } {\lambda}^r\brk*{y} A_{r,\beta}\brk*{\hat{x}} - \mathbb{E}_{p_{\beta}\brk*{\hx}}\brk[s]*{{\lambda}^{0}_{\beta}\brk*{\hx} } \\
        &=  H\brk*{Y} - \mathbb{E}_{p_{\beta}\brk*{\hx}}\brk[s]*{ \sum_{r=1}^{d}{\lambda}_{\beta}^r\brk*{\hat{x}}A_{r,\beta}\brk*{\hat{x}} + {\lambda}^{0}_{\beta}\brk*{\hx} }
\end{align*}

\section{Optimizing the error exponent}
\label{app:err_exp}
We start by to expressing the Chernoff information for the binary hypothesis testing problem using $p\brk*{y
\mid x}$: 
\begin{align*}
C\brk*{p_{0}, p_1} &= \min_{\lambda \in \brk[s]*{0,1}} \log \brk*{\sum_{x} p\brk*{x \mid y_0}^{q_{\lambda}\brk*{y_0}} p\brk*{x \mid y_1}^{q_{\lambda}\brk*{y_1}} } \\
&= \min_{\lambda \in \brk[s]*{0,1}} \log \brk*{\sum_{x} p\brk*{y=0 \mid x}^{\lambda}p\brk*{x}^{\lambda} p\brk*{y=0}^{-\lambda}p\brk*{y=1 \mid x}^{1-\lambda}p\brk*{x}^{1-\lambda} p\brk*{y=1}^{\lambda-1}} \nonumber \\
&= \min_{\lambda \in \brk[s]*{0,1}} \log \brk*{\sum_{x}p\brk*{x} p\brk*{y=0 \mid x}^{\lambda} p\brk*{y=1 \mid x}^{1-\lambda} } -\log\brk*{p\brk*{y=0}^{\lambda}p\brk*{y=1}^{1-\lambda}} \nonumber \\
&= \min_{q_{\lambda}\brk*{y}} \log \brk*{\sum_{x} e^{q_{\lambda}\brk*{y_0} \log p\brk*{x \mid y_0} + q_{\lambda}\brk*{y_1}\log p\brk*{x \mid y_1}} } \\
&=\min_{q_{\lambda}\brk*{y}} \log \brk*{\sum_{x}  e^{-D\brk[s]*{q_{\lambda}\brk*{y} \| p\brk*{y \mid x}} + D\brk[s]*{q_{\lambda}\brk*{y} \| p{y}} + \log p\brk*{x}} }\\
&= \min_{q_{\lambda}\brk*{y}} \log \brk*{e^{D\brk[s]*{q_{\lambda}\brk*{y} \| p\brk*{y}}} \sum_{x}  e^{-D\brk[s]*{q_{\lambda}\brk*{y} \| p\brk*{y \mid x}}  + \log p\brk*{x}} } \\
&= \min_{q_{\lambda}\brk*{y}} \brk[c]*{ \log \brk*{ \sum_{x} p\brk*{x}e^{-D\brk[s]*{q_{\lambda}\brk*{y} \| p\brk*{y \mid x}}} } + D\brk[s]*{q_{\lambda}\brk*{y} \| p\brk*{y}}}
,\end{align*}
where $q_{\lambda}\brk*{y_0} = \lambda, q_{\lambda}\brk*{y_1} = 1-\lambda$.
Now, if we consider the mapping, $q_{\lambda}\brk*{y} =  p_{\beta}\brk*{y \mid \hx}$ we can write the above as:
\begin{align*}
C\brk*{p_{0}, p_1} &= \min_{p_{\beta}\brk*{y \mid \hx}}\brk[c]*{ \log \brk*{\sum_{x} p\brk*{x} e^{-D\brk[s]*{p_{\beta}\brk*{y \mid \hx}\| p\brk*{y \mid x}}} } + D\brk[s]*{ p_{\beta}\brk*{y \mid \hx} \| p\brk*{y}} }
.\end{align*}
The above term in minimization is proportional to $\log$-partition function of $p_{\beta}\brk*{x \mid \hx}$, namely we get the mapping $p_{\beta}\brk*{x \mid \hx} = p_{\lambda}$. 
Next we shall generalize the setting to the $M$-hypothesis testing problem. Having that solving for the Chernoff information is notoriously difficult we consider an upper bound to it, taking the expectation over the classes. Instead of  choosing $p_{\lambda^*}$ as the maximal value of the minimimum  $\brk[c]*{D\brk[s]*{p_{\lambda^*} \| p_0}, D\brk[s]*{p_{\lambda^*} \| p_1}}$ we consider it w.r.t the full set $\brk[c]*{D\brk[s]*{p_{\lambda^*} \| p_i}}_{i=1}^{M}$. Using the above mapping we must take the expectation also over the representation variable $\hx$. Thus we get the expression:
\begin{align*} \label{eq:chernoff_dual}
    D^{*}\brk*{\beta} &= \min_{p_{\beta}\brk*{y\mid \hx}, p_{\beta}\brk*{ \hx \mid x}}\mathbb{E}_{p_{\beta}\brk*{y, \hx}}\brk[s]*{D\brk[s]*{ p_{\beta}\brk*{ x \mid \hx} \mid p\brk*{x\mid y}}}
.\end{align*}
  From the definition of $D^{*}\brk*{\beta}$ we obtain the desired bound of the {\dualib}:
  \begin{align*} 
    D^{*}\brk*{\beta} &= \min_{p_{\beta}\brk*{y\mid \hx}, p_{\beta}\brk*{ \hx \mid x}}\mathbb{E}_{p_{\beta}\brk*{y, \hx}}\brk[s]*{D\brk[s]*{ p_{\beta}\brk*{ x \mid \hx} \| p\brk*{x\mid y}}} \\
    &= \min_{p_{\beta}\brk*{y\mid \hx}, p_{\beta}\brk*{ \hx \mid x}}\sum_{x, y,\hx} p_{\beta}\brk*{y \mid\hx}p_{\beta}\brk*{\hx} \brk[s]*{D\brk[s]*{ p_{\beta}\brk*{ x \mid \hx} \mid p\brk*{x\mid y}}} \\
    &= \min_{p_{\beta}\brk*{y\mid \hx}, p_{\beta}\brk*{ \hx \mid x}}\sum_{x, y,\hx} p_{\beta}\brk*{y \mid\hx}p_{\beta}\brk*{\hx}p_{\beta}\brk*{x \mid\hx} \brk[c]*{ \log \frac {p_{\beta}\brk*{y \mid\hx}}{p\brk*{y\mid x}} + \log \frac {p_{\beta}\brk*{x \mid\hx}}{p_{\beta}\brk*{y \mid\hx}} + \log \frac{p\brk*{y}}{p\brk*{x}} } \\
    &=  \min_{p_{\beta}\brk*{y\mid \hx}, p_{\beta}\brk*{ \hx \mid x}}\brk[c]*{I(X;\hX) + \mathbb{E}_{p_{\beta}\brk*{x, \hx}}\brk[s]*{D\brk[s]*{ p_{\beta}\brk*{ y \mid \hx} \| p\brk*{y \mid x}}} + H(Y \mid \hX) + \mathbb{E}_{p_{\beta}\brk*{y}}\brk[s]*{\log p\brk*{y}} }\\
    &\leq \min_{p_{\beta}\brk*{y\mid \hx}, p_{\beta}\brk*{ \hx \mid x}}\brk[c]*{I(X;\hX) + \mathbb{E}_{p_{\beta}\brk*{x, \hx}}\brk[s]*{D\brk[s]*{ p_{\beta}\brk*{ y \mid \hx} \| p\brk*{y \mid x}}}} \\
    &\leq \mathcal{F}^{*}\brk[s]*{p\brk*{\hat{x} \mid x}; p\brk*{y \mid \hat{x} }}
.\end{align*}
  
  \subsection{Error exponent optimization example}
\begin{figure}[htb!]
    \begin{center}
        \includegraphics[scale = 0.42]{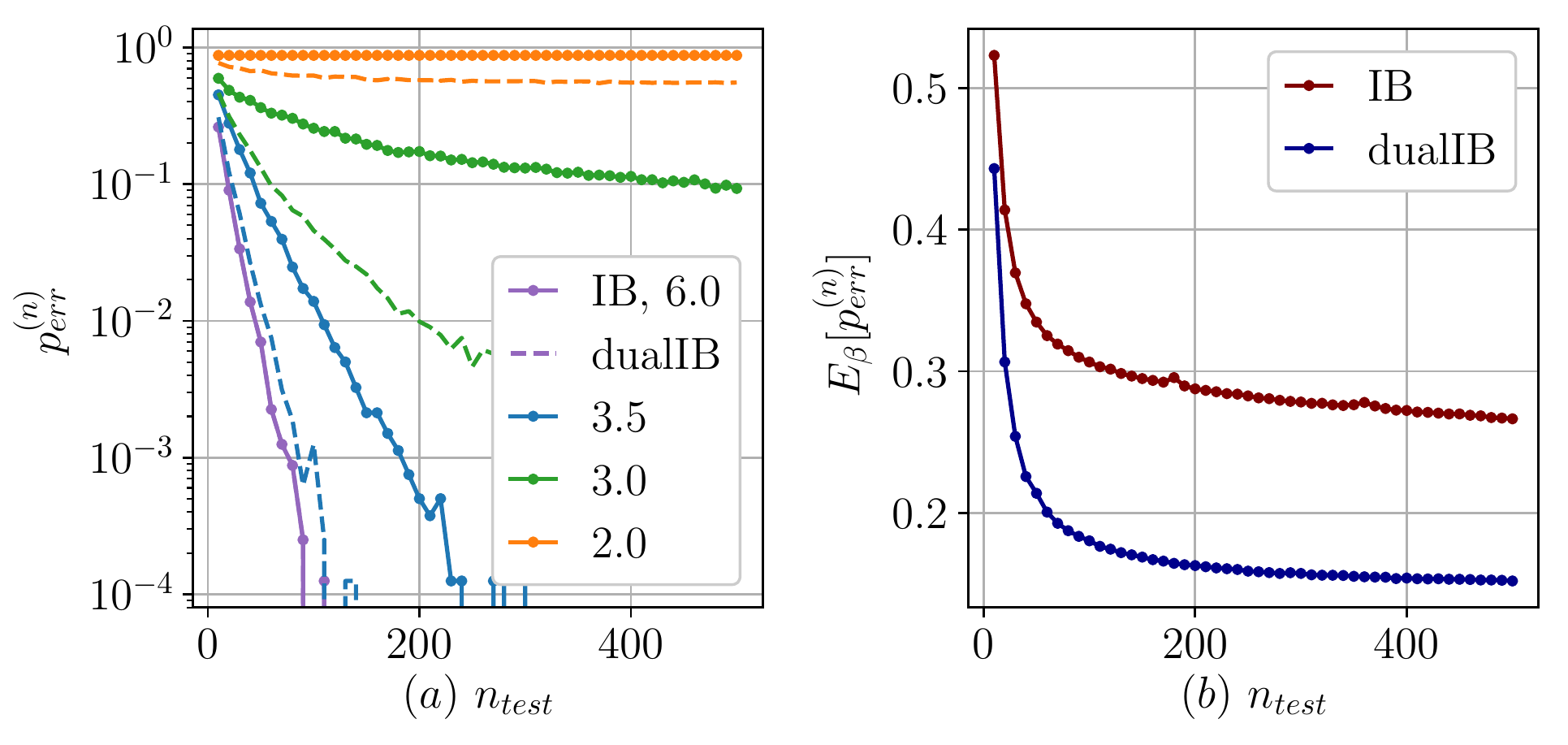}
    \end{center}
        
        \caption{The probability of error, $p_{err}^{\brk*{n}}$, as a function of test sample size, $n_{test}$. $\brk*{a}$ The exponential decay of error for representative $\beta$ values ($\log_2(\beta)$ reported in the legend). For a given $\beta$ the {\ib} performance is plotted in solid line and the {\dualib} in dashed (for $\log_{2}\brk*{\beta} = 6$ the lines overlap). $\brk*{b}$ The expectation of the error over all $\beta$'s ($\log_2\beta \in \brk[s]*{1, 6}$).}
        
        \label{fig:p_err}
  \end{figure}

To demonstrate the above properties we consider a classification problem with $M=8$ classes, each class characterized by $p_{i} = p\brk*{x \mid y_i}$. The \emph{training} is performed according to the above algorithms to obtain the {\ib} ({\dualib}) encoder and decoder. For the prediction, given a new sample $x^{\brk*{n}} \overset{i.i.d}{\sim} p\brk*{x\mid y}$ defining an empirical distribution $\hat{p}\brk*{x}$ the prediction is done by first evaluating $\hat{p}_{\beta}\brk*{\hx} = \sum_{x} p_{\beta}\brk*{\hx \mid x}\hat{p}\brk*{x}$. Next, using the (representation) optimal decision rule, we obtain the prediction:
\begin{align*} 
    \hat{H}_{\beta} = \arg \min_{i} D\brk[s]*{\hat{p}_{\beta}\brk*{\hx} \| p_{\beta}\brk*{\hx \mid y_i}  }  
    ,\end{align*}
and we report $p_{err}^{\brk*{n}}$, the probability of miss-classification. 
This represents the most general classification task; the distributions $p_i$ represent the empirical distributions over a training data-set and then testing is performed relative to a test set.
Looking at the results, \fref{fig:p_err}, it is evident that indeed the {\dualib} improves the prediction error (at $log_2\brk*{\beta} = 6$ the algorithms performance is identical due to the similarity of the algorithms behavior as $\beta$ increases).

\section{The variational {\dualib} } 

\subsection{Derivation of the {\vdib} objective}
\label{app:vdib_obj}
Just as \citep{fischer2020ceb} did, we can variationally upper bound the information of the input with the representation variable using:
\begin{align*}
    I(\hX;X \mid Y)=\mathbb{E}_{p(x,y)p(\hx\mid x)}\left[\log\frac{p(\hx \mid x,y)}{p(\hat{x}\mid y)}\right]\leq \mathbb{E}_{\tilde{p}(y\mid x) p\brk*{x}p(\hat{x}\mid x)}\left[\log\frac{p(\hat{x}\mid x)}{q(\hat{x}\mid y)}\right]
\end{align*}
where $q(\hx \mid y)$ is a variational class conditional marginal.
In contradiction to the CEB, in order to bound the {\dualib} distortion, we replace the bound on $I(\hX;Y)$ with a bound over the expected {\dualib} distortion. Here, given the assumption of a noise model $\tilde{p}\brk*{y \mid x}$ which we evaluate the expected distortion with respect to it:
\begin{align*}
    \mathbb{E}_{p\brk*{x, \hx}}\brk[s]*{d_{\dualib}\brk*{x, \hx}} &=  \mathbb{E}_{{p}\brk*{y\mid\hx}p\brk*{\hx\mid x} p\brk*{x}}\brk[s]*{\log \frac{p\brk*{y\mid \hx}}{\tilde{p}\brk*{y\mid x}}} 
\end{align*}
Combining the above together gives the variational upper bound to the {\dualib} as the following objective:
\begin{align*}
    I(X;\hat{X}) + \beta \mathbb{E}_{p\brk*{x, \hx}}\brk[s]*{d_{\dualib}\brk*{x, \hx}}  \leq   
   {\mathbb{E}_{\tilde{p}(y\mid x)p\brk*{\hx\mid x} p\brk*{x}}\brk[s]*{
   \log\frac{p\brk*{\hx\mid x}}{q\brk*{\hx\mid y}} } + \beta \mathbb{E}_{{p}\brk*{y\mid\hx}p\brk*{\hx\mid x}p\brk*{x}}\brk[s]*{\log \frac{p\brk*{y\mid \hx}}{\tilde{p}\brk*{y\mid x}}}}
\end{align*}
\subsection{Experimental setup}
\label{app:exp_setup}
For  both CIFAR10 and FasionMNIST We trained a set of 30 $28-10$ Wide ResNet
models in a range of values of $\beta$ ($-5\leq \log\beta \leq 5$). The training was doneusing Adam \citep{kingma2014adam} at a base learning rate of $10^{-4}$. We lowered the learning rate two times by a factor of $0.3$ each time.
Additionally,  following \cite{fischer2020ceb}, we use a jump-start method for $\beta<100$. We start the training with $\beta=100$, anneal down  to the target $\beta$ over 1000 steps. The training includes data augmentation with horizontal flip and width height shifts. Note, that we exclude from the analysis runs that  didn't succeed to learn at all (for which the results look as random points).

\subsection{The variational information plane}
Note that, for the information plane analysis, there were several runs that failed to achieved more than random accuracy. In such cases, we remove them. 
The confusion matrix used for the FashionMNIST data-set is:
\begin{align*}
\begin{pmatrix}
0.828 & 0.013 & 0.012 & 0.011 & 0.018 & 0.& 0.002 & 0.004 & 0.085 & 0.027  \\
0.01&0.91&0.&0.005&0.001&0.001&0.&0.001&0.011&0.061  \\
0.047&0.001&0.708&0.064&0.088&0.014&0.063&0.004&0.008&0.003 \\
0.003&0.004&0.016&0.768&0.033&0.093&0.05&0.019&0.004&0.01 \\
0.01&0.&0.039&0.043&0.788&0.012&0.057&0.043&0.006&0.002 \\
0.002&0.&0.01&0.137&0.029&0.777&0.008&0.033&0.&0.004 \\
0.007&0.002&0.01&0.054&0.029&0.007&0.888&0.001&0.001&0.001 \\
0.024&0.002&0.014&0.039&0.076&0.017&0.004&0.818&0.002&0.004 \\
0.027&0.013&0.&0.007&0.003&0.&0.003&0.&0.933&0.014. \\
0.019&0.064&0.001&0.007&0.002&0.001&0.001&0.&0.018&0.887
\end{pmatrix}    
\end{align*}

\subsection{The {\vdib} noise models}
\label{app:noise_model}
\begin{figure}[htb!]
    \centering
    \includegraphics[width=\textwidth]{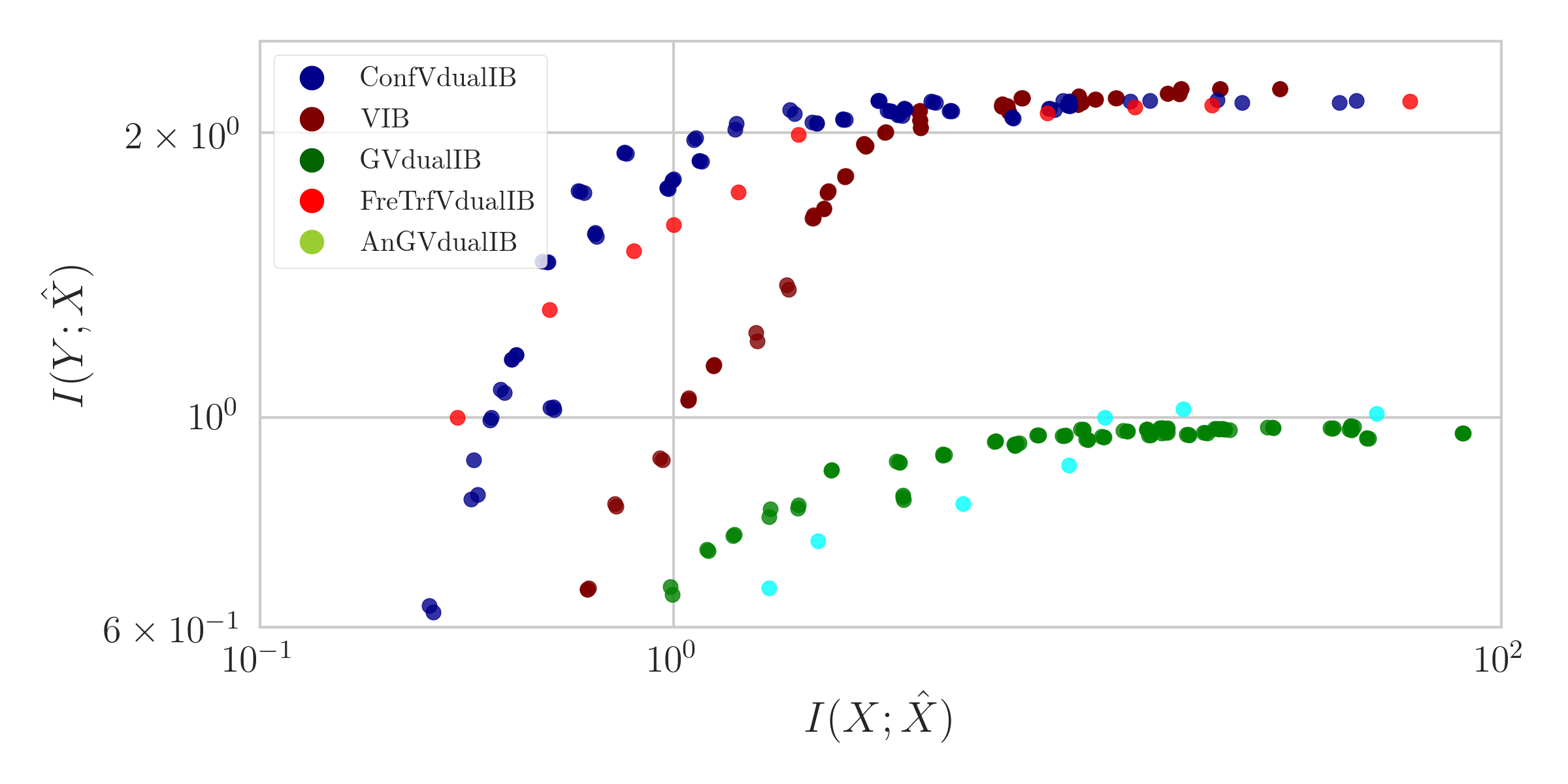}
 \caption{The information plane for the different noise models.}
 \label{fig:noise_models}
\end{figure}

As described in the main text we consider two additional noise models; (i) An analytic Gaussian integration of the log-loss around the one-hot labels ({\anvdib}) (ii) 
Using predictions of another trained model as the induced distribution ({\prdvib}). In this case, we use a deterministic wide ResNet $28-10$ network that achieved $95.8\%$ accuracy on CIFAR10. In \fref{fig:noise_models} we can see all the different models, $4$ noise models for the {\vdib} and the {\vib}). As expected, we can see that analytic Gaussian integration  noise model obtains similar results to adding Gaussian noise to the one-hot vector of the true label, while the performance of the noise models that are based on a trained network are similar to the {\cvdib}. 

\subsubsection{Training {\vib} model with noise}
\begin{figure}[htb!]
    \centering
    \includegraphics[width=\textwidth]{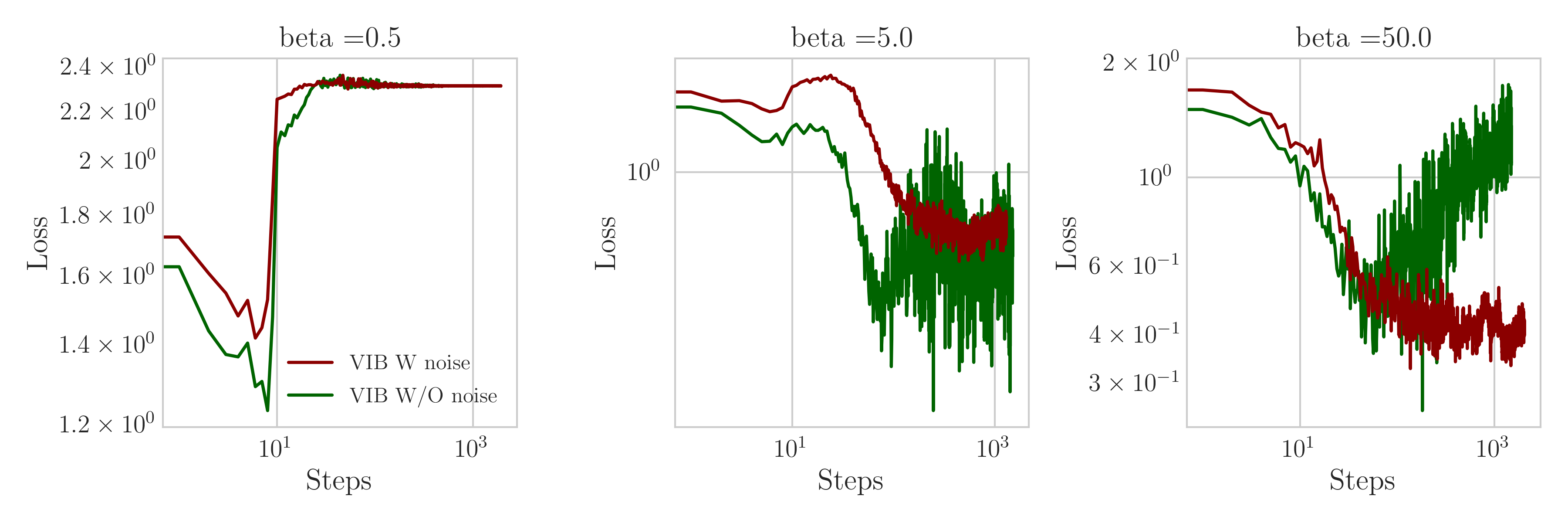}
 \caption{The influence of a noise model on the {\vib} performance. Loss as function of the update steps for different values of $\beta$, $\beta = 0.5, 5.0. 50.0$ from left to right.}
 \label{fig:label_smooth}
\end{figure}

In our analysis, we train a {\vib} model with the same noise model as the {\vdib}. Namely, instead of training with a deterministic label (one-hot vector of zeros and ones), we use our noise model also for the {\vib}. As mentioned in the text, this training procedure is closely related to label smoothing. In \fref{fig:label_smooth}, we present the loss function of the {\vib} on CIFAR10 with and without the noise models along the training process for $3$  different values of $\beta$. For a small $\beta$ (left) both regimes under-fit the data as expected. However, when we enlarge $\beta$, we can see that the labels' noise makes the training more stable and for a high value of $\beta$ (right) training without noise over-fits the data and the loss increases.

The confusion matrix for the CIFAR10 data set is:
\begin{align*}
  \begin{pmatrix}
0.878 & 0.  & 0.017& 0.013&0.002& 0.001& 0.082& 0.   & 0.007&
        0.   \\
       0.  & 0.984& 0.002&0.009& 0.001& 0.   & 0.003& 0.   & 0.   &
        0.   \\
       0.013& 0.001& 0.896& 0.009& 0.038& 0.   & 0.043& 0.   & 0.   &
        0.   \\
       [0.022& 0.004& 0.011& 0.913& 0.023& 0.   & 0.027& 0.   & 0.001&
        0.   \\
       0.   & 0.   & 0.072& 0.022& 0.85 & 0.   & 0.058& 0.   & 0.   &
        0.\\   
       0.   & 0.   & 0.  & 0.   & 0.   & 0.982& 0.   & 0.011& 0.   &
        0.007\\
       0.099& 0.001& 0.049& 0.021& 0.055& 0.   & 0.768& 0.   & 0.005&
        0.   \\
       0.   & 0.   & 0.   & 0.   & 0.   & 0.006& 0.   & 0.976& 0.   &
        0.019\\
       0.004& 0.001& 0.001& 0.001& 0.004&0.002&0.003& 0.001& 0.98 &
        0.001\\
       0.   & 0.   & 0.   & 0.   & 0.   & 0.004& 0.   & 0.02 & 0.001&
        0.974
\end{pmatrix}  
\end{align*}

\subsection{CIFAR100 results}
\label{app:cifar100}
\begin{figure}[htb!]
\begin{subfigure}{.5\textwidth}
    \centering
    \includegraphics[width=\textwidth]{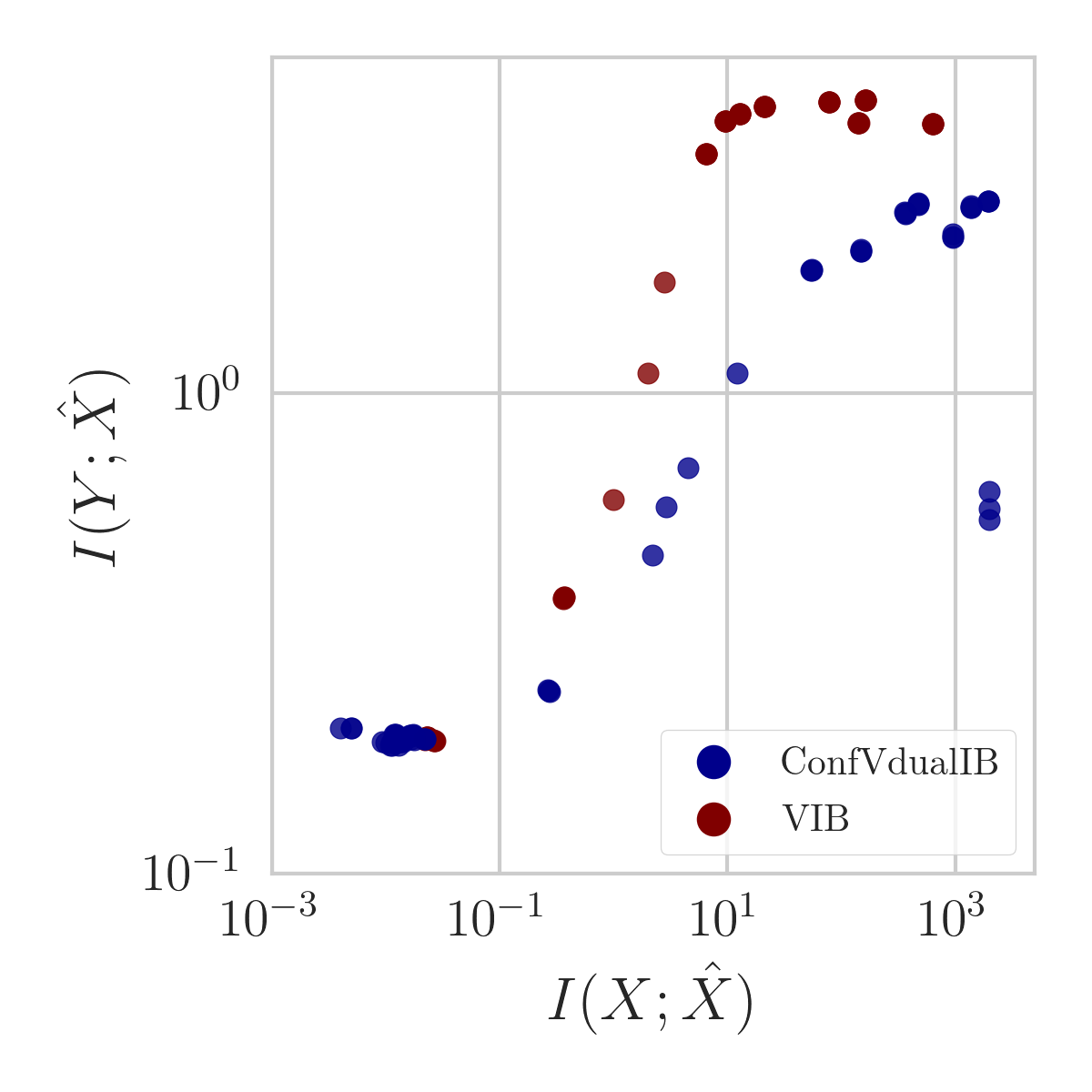}
    \caption{The information plane}
    \label{fig:inf_plane_cifar100_beta}
\end{subfigure}
\begin{subfigure}{.5\textwidth}
    \centering
    \includegraphics[width=\textwidth]{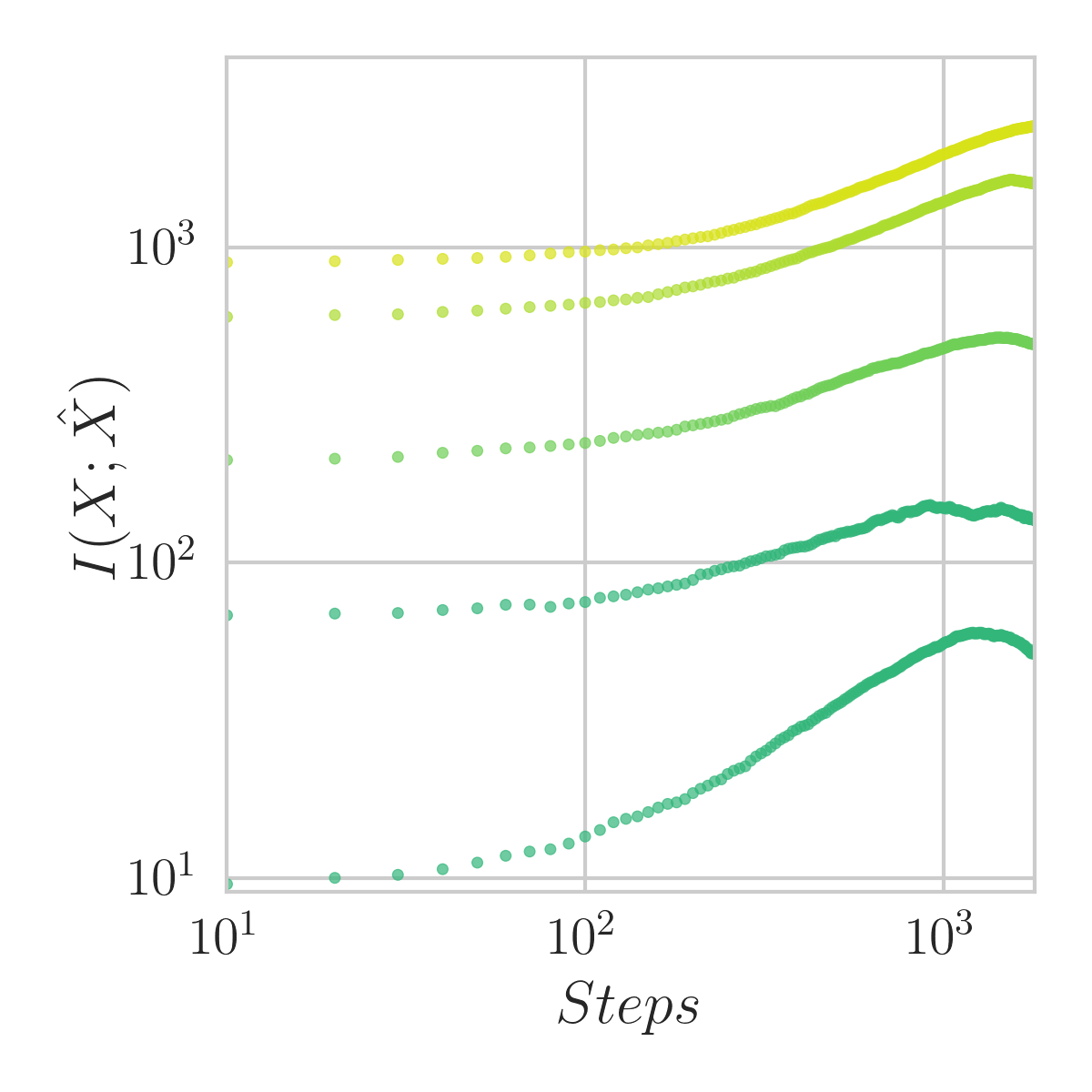}
    \caption{$I(X;\hX)$ vs. update steps for different $beta$'s}
    \label{fig:inf_plane_cifar100_steps}
\end{subfigure}
 \caption{Experiments over CIFAR100. $(a)$ The information plane of the {\cvdib} and {\vib} for a range of $\beta$ values at the final training step. $(b)$ The evolution of the the {\cvdib}'s $I(X;\hX)$ along the optimization update steps.}
 \label{fig:inf_plane_cifar100}
\end{figure}

As mentioned in the text, we trained {\vdib} networks also on CIFAR100. For this, we used the  same  $28-10$ Wide ResNet with a confusion matrix as our noise model.  The confusion matrix was calculated based on the predictions of a deterministic network. The deterministic network achieved $80.2\%$ accuracy on CIFAR100.
In \ref{fig:inf_plane_cifar100_beta}, we can see the information plane for both {\vdib} and the {\vib} models. As we can see, both models are monotonic with $I(X;\hX)$, however the {\vib}'s performance is better. The {\vib} achieves higher values of information with the labels along with more compressed representation at any given level of predication. Although a broader analysis is required, and possible further parameter tuning of the architecture, we hypothesize that the caveat is in the noise model used for the {\vdib}. Using a noise model which is based on a network that achieves almost $20\%$ error might be insufficient in this case. It might be that ``errors'' in the noise model becomes similar to random errors, similar to the Gaussian case, and hence depicting similar learning performance to the {\gvdib} case. 

When we look at the information with the input as a function of time ,\fref{fig:inf_plane_cifar100_steps}, we see that similar to the FasionMNIST and CIFAR10 results, the information saturates for small values of $\beta$, but over-fits for higher values of it. 

\end{document}